\documentclass[fleqn,10pt]{article}
\usepackage{latexsym, graphicx, epsfig, amsmath, amssymb,amsfonts}
\usepackage{natbib,amsthm,version}
\usepackage{amsbsy,bm,multirow,enumerate}
\usepackage[titletoc,page]{appendix}
\usepackage[mathscr]{eucal}
\usepackage{mathtools}
\usepackage{color}
\usepackage[utf8]{inputenc}
\usepackage[english]{babel}
\usepackage{amsthm}
\usepackage{enumerate}

\newtheorem{theorem}{Theorem}[section]

\theoremstyle{definition}

\title{
  Numerical Approximation of Incompressible Navier-Stokes Equations
  Based on an Auxiliary Energy Variable
} 
\author{
  Lianlei Lin$^{1,2}$, \
  Suchuan Dong$^1$\thanks{Author of correspondence. Email: sdong@purdue.edu} \\
  $^1$Center for Computational and Applied Mathematics \\
  Department of Mathematics \\
  Purdue University, USA \\
  $^2$School of Electrical Engineering and Automation\\
  Harbin Institute of Technology, China 
 } 

\date{} 
\begin{document}
\maketitle



\begin{abstract}

We present a numerical scheme for approximating the incompressible
Navier-Stokes equations based on an auxiliary variable associated with
the total system energy. By introducing a dynamic equation for
the auxiliary variable and reformulating the Navier-Stokes equations
into an equivalent system, the scheme satisfies a discrete
energy stability property
in terms of a modified energy and it allows for an efficient 
solution algorithm and implementation. Within each time step,
the algorithm involves the computations of
two pressure fields and two velocity fields by solving
several de-coupled individual 
linear algebraic systems with constant coefficient matrices,
together with the solution of a nonlinear algebraic equation
about a {\em scalar number} involving a negligible cost.
A number of numerical experiments are presented to demonstrate the accuracy
and the performance of the presented algorithm.

\end{abstract}


\vspace{0.05cm}
Keywords: {\em 
  energy stability;
  unconditional stability;
  Navier-Stokes equations;
  incompressible flows;
  auxiliary variable
}

\section{Introduction}
\label{sec:intro}

%
%


We focus on the numerical approximation of the incompressible
Navier-Stokes equations, which form the basis for simulations of
single-phase incompressible flows and turbulence~\cite{KravchenkoM2000,MaKK2000,DongKER2006,Dong2007}.
The momentum equations for two-phase and multiphase
flow problems~\cite{YueFLS2004,AbelsGG2012,DongS2012,Dong2018}
often take a form analogous to
the incompressible Navier-Stokes equations with a similar
mathematical structure. Developing efficient and effective algorithms
for incompressible Navier-Stokes equations therefore can have implications
in fields far beyond incompressible flows.


An essential tradeoff confronting a practitioner of computational fluid dynamics (CFD) is
between the desire to be able
to use larger time step sizes (permitted by accuracy/stability)
and the computational cost.
On one end of the spectrum, unconditionally energy-stable
schemes~(see e.g.~\cite{Shen1992,SimoA1994,LabovskyLMNR2009,DongS2010,Sanderse2013,JiangMRT2016,ChenSZ2018}, among others)
can alleviate the time step size constraint,
and one can concentrate on the accuracy requirement when choosing
a time step size in a simulation task.
This however comes with a downside.
Within each time step energy-stable schemes typically
involve the solution  of a system of nonlinear
algebraic equations or a system of linear
algebraic equations with a variable and time-dependent coefficient matrix,
inducing a high computation cost due to the Newton nonlinear iterations
and the need for frequent re-computations of the coefficient matrices.
This can render the overall approach 
inefficient, and therefore such schemes are not often
used in large-scale production simulations in practice,
especially for dynamic problems.
%
On the other end of the spectrum, 
semi-implicit splitting  type (or fractional-step)
schemes (see e.g.~\cite{Chorin1968,Temam1969,KimM1985,KarniadakisIO1991,BrownCM2001,XuP2001,GuermondS2003,LiuLP2007,HyoungsuK2011,SersonMS2016},
the review article
\cite{GuermondMS2006}, and the references therein)
de-couple the computations for the velocity and the pressure fields,
and all the coefficient matrices are constant and can be pre-computed.
Such schemes have a low computational cost. But they are only conditionally
stable, and the time step size
that can be used is restricted by e.g.~CFL type conditions.
Thanks to their low cost, semi-implicit schemes are very popular 
in large-scale production simulations for flow physics
investigations (see e.g.~\cite{MaKK2000,DongKER2006,Dong2009,DongZ2011}).

We present in this work a numerical scheme
for the incompressible Navier-Stokes equations that in a sense 
resides somewhere between the two extremes on the spectrum of methods.
The salient feature of this scheme is the introduction of an auxiliary
variable (which is a scalar number) associated with the
total energy of the Navier-Stokes
system. This idea is inspired by a recent work~\cite{ShenXY2018}
for gradient-type dynamical systems. The incompressible Navier-Stokes
equations are then reformulated into an equivalent system employing
the auxiliary energy variable, and a numerical scheme is devised to approximate
the reformulated system.
Within a time step, the algorithm requires the computations
of two pressure fields and two velocity fields, as well as
the solution of a nonlinear algebraic equation about a scalar number.
The computation for each of the pressure/velocity fields involves a linear
algebraic system with a constant coefficient matrix that can be
pre-computed.
Solving the nonlinear algebraic equation requires Newton iterations,
but its cost is negligible (accounting for $1\sim 2\%$ of the total cost per
time step) because this nonlinear equation is about {\em a scalar number},
not a field function.
This scheme can be shown to satisfy a discrete energy
stability property in terms of
a modified energy. Numerical experiments show that the algorithm allows
the use of very large time step sizes for steady flow problems, and for unsteady
flow problems it appears to be not as effective in terms of the time step
size. The amount of operations (and the computational cost) of this
algorithm per time step is approximately twice that of the semi-implicit scheme
from~\cite{Dong2015clesobc}, which is only conditionally stable.


The novelties of this work lie in several aspects:
(i) the introduction of the auxiliary energy variable into
and the resultant reformulation of the Navier-Stokes system;
(ii) the numerical scheme for approximating the reformulated
system of equations; and
(iii) the efficient solution procedure for overcoming
the difficulty caused by the unknown auxiliary variable in the implementation of
the scheme.


The rest of this paper is structured as follows.
In Section \ref{sec:method} we introduce an auxiliary energy variable
and reformulate the Navier-Stokes equations based on this variable.
A numerical scheme is presented for approximating the reformulated equivalent
system. We then discuss how to implement the scheme, and in particular
how to overcome the challenge caused by the unknown auxiliary variable
in the implementation.
In Section \ref{sec:tests} we present several representative
numerical simulations to test the accuracy and performance of the algorithm.
Section \ref{sec:summary} concludes the presentation with
some closing remarks.

\section{Auxiliary Variable-Based Algorithm for Incompressible Navier-Stokes Equations}
\label{sec:method}


\subsection{Reformulated Equations and Numerical Scheme Formulation}

Consider an incompressible flow contained in some
domain $\Omega$ in two or three dimensions, whose
boundary is denoted by $\partial\Omega$.
The dynamics of the flow is described by
the incompressible Navier-Stokes equations,
given in a non-dimensional form as follows,
\begin{subequations}
\begin{equation}
  \frac{\partial\mathbf{u}}{\partial t}
  + \mathbf{N}(\mathbf{u}) 
  + \nabla p
  - \nu\nabla^2\mathbf{u}
  = \mathbf{f},
  \label{equ:nse}
\end{equation}
\begin{equation}
  \nabla\cdot\mathbf{u} = 0,
  \label{equ:continuity}
\end{equation}
\end{subequations}
where $\mathbf{x}$ and $t$ denote the spatial
coordinate and time, $\mathbf{u}(\mathbf{x},t)$
and $p(\mathbf{x},t)$ are respectively the normalized velocity and
pressure, $\mathbf{f}(\mathbf{x},t)$
is an external body force, and $\mathbf{N}(\mathbf{u})$
is the convection term,
$\mathbf{N}(\mathbf{u})=\mathbf{u}\cdot\nabla\mathbf{u}$.
$\nu$ denotes the inverse of the Reynolds number $Re$,
\begin{equation}
  \nu = \frac{1}{Re} = \frac{\nu_f}{U_0L}
  \label{equ:def_Re}
\end{equation}
where $U_0$ is the characteristic velocity scale, $L$ is the
characteristic length scale,
and $\nu_f$ is the kinematic viscosity of the fluid.
On the domain boundary $\partial\Omega$ we assume
that the velocity is known
\begin{equation}
  \mathbf{u} = \mathbf{w}(\mathbf{x},t),
  \quad \text{on} \ \partial\Omega
  \label{equ:bc_vel}
\end{equation}
where $\mathbf{w}$ denotes the boundary velocity.
The system is supplemented by the initial condition
\begin{equation}
  \mathbf{u}(\mathbf{x},0) = \mathbf{u}_{in}(\mathbf{x})
  \label{equ:ic_vel}
\end{equation}
where the initial velocity distribution $\mathbf{u}_{in}$
is assumed to be compatible with the boundary
condition \eqref{equ:bc_vel} on $\partial\Omega$ and
satisfies the equation \eqref{equ:continuity}.
In the governing equations \eqref{equ:nse}--\eqref{equ:continuity},
only the pressure gradient $\nabla p$ is physically meaningful,
and the absolute pressure value is not fixed (pressure can be
shifted by an arbitrary constant).
In order to fix the pressure values in the numerical solution
we impose the following often-used condition
\begin{equation}
  \int_{\Omega} p = 0.
  \label{equ:p_condition}
\end{equation}


Consider a shifted total energy of the system
\begin{equation}
E(t) = E[\mathbf{u}] = C_0 + \int_{\Omega} \frac{1}{2}|\mathbf{u}|^2 
\label{equ:def_energy}
\end{equation}
where $C_0$ is a chosen constant such that $E(t)>0$ for all
$t\geqslant 0$.
Define an auxiliary variable $R(t)$ by
\begin{equation}
  R(t) = \sqrt{E(t)}.
  \label{equ:def_R}
\end{equation}
Then
\begin{equation}
  2R\frac{dR}{dt} = \frac{dE}{dt}
  = \int_{\Omega} \frac{\partial\mathbf{u}}{\partial t}\cdot\mathbf{u}
  = \int_{\Omega}\left(
  \frac{\partial\mathbf{u}}{\partial t}
  + \mathbf{u}\cdot\nabla\mathbf{u}
  \right) \cdot \mathbf{u}
  -\int_{\partial\Omega}(\mathbf{n}\cdot\mathbf{u})\frac{1}{2}|\mathbf{u}|^2
  \label{equ:R_equ}
\end{equation}
where $\mathbf{n}$ is the outward-pointing unit vector
normal to the boundary $\partial\Omega$, and
we have used integration by part, the equation \eqref{equ:continuity},
and the divergence theorem.
It should be emphasized that both $R(t)$ and $E(t)$ are scalar
variables, not field functions. At $t=0$,
\begin{equation}
  R(0) = \left( C_0 +
  \int_{\Omega} \frac{1}{2}|\mathbf{u}_{in}|^2
  \right)^{\frac{1}{2}}.
  \label{equ:ic_R}
\end{equation}

In light of the definition \eqref{equ:def_R},
we re-write equation \eqref{equ:nse} into an
equivalent form
\begin{equation}
  \frac{\partial\mathbf{u}}{\partial t}
  +\frac{R(t)}{\sqrt{E(t)}}\mathbf{N}(\mathbf{u})
  + \nabla p
  - \nu\nabla^2\mathbf{u}
  = \mathbf{f}.
  \label{equ:nse_mod}
\end{equation}
We also re-write equation \eqref{equ:R_equ} into
an equivalent form
\begin{equation}
   2R\frac{dR}{dt} =
    \int_{\Omega}\left[
  \frac{\partial\mathbf{u}}{\partial t}
  + \frac{R(t)}{\sqrt{E(t)}}\mathbf{N}(\mathbf{u})
  \right] \cdot \mathbf{u}
  -  \int_{\partial\Omega}(\mathbf{n}\cdot\mathbf{u})\frac{1}{2}|\mathbf{u}|^2.
  \label{equ:R_equ_mod}
\end{equation}

The original system consisting of the equations
\eqref{equ:nse}--\eqref{equ:continuity},
\eqref{equ:bc_vel} and \eqref{equ:ic_vel}
is equivalent to the reformulated system
consisting of equations \eqref{equ:nse_mod},
\eqref{equ:continuity}, \eqref{equ:R_equ_mod},
together with the boundary condition \eqref{equ:bc_vel}
and the initial conditions \eqref{equ:ic_vel}
and \eqref{equ:ic_R},
in which $E(t)$ is given by \eqref{equ:def_energy}.
We next focus on this reformulated equivalent system of
equations, and present a numerical scheme
for approximating this system. 


Let $n\geqslant 0$ denote the time step index, and
$(\cdot)^n$ denote the variable $(\cdot)$ at
time step $n$. Let $J$ ($J=1$ or $2$) denote the
temporal order of accuracy of the scheme.
We set
\begin{equation}
  \mathbf{u}^0 = u_{in}(\mathbf{x}), \quad
  R^0 = \sqrt{E(0)} = \sqrt{C_0 + \int_{\Omega} \frac{1}{2}|\mathbf{u}^0|^2}.
\end{equation}
Then given ($\mathbf{u}^n$, $p^n$, $R^n$), we compute
($\mathbf{u}^{n+1}$, $p^{n+1}$, $R^{n+1}$) through
the following scheme
\begin{subequations}
  \begin{equation}
    \frac{\gamma_0\mathbf{u}^{n+1}-\hat{\mathbf{u}}}{\Delta t}
    + \frac{R^{n+1}}{\sqrt{E^{n+1}}}\mathbf{N}(\bar{\mathbf{u}}^{n+1})
    + \nabla p^{n+1}
    - \nu\nabla^2\mathbf{u}^{n+1}
    = \mathbf{f}^{n+1},
    \label{equ:alg_1}
  \end{equation}
  \begin{equation}
    \nabla\cdot\mathbf{u}^{n+1} = 0,
    \label{equ:alg_2}
  \end{equation}
  \begin{equation}
    \begin{split}
    2R^{n+1}\frac{\gamma_0R^{n+1} - \hat{R}}{\Delta t}
    =&
     \int_{\Omega}\left[
      \frac{\gamma_0\mathbf{u}^{n+1}-\hat{\mathbf{u}}}{\Delta t}
      + \frac{R^{n+1}}{\sqrt{E^{n+1}}}\mathbf{N}(\bar{\mathbf{u}}^{n+1})
      \right]\cdot\mathbf{u}^{n+1} 
    - \int_{\partial\Omega}(\mathbf{n}\cdot\mathbf{u}^{n+1})\frac{1}{2}|\mathbf{u}^{n+1}|^2,
    \end{split}
    \label{equ:alg_3}
  \end{equation}
  \begin{equation}
    \mathbf{u}^{n+1} = \mathbf{w}^{n+1},
    \quad \text{on} \ \partial\Omega.
    \label{equ:alg_4}
  \end{equation}
  \begin{equation}
    E^{n+1} = \int_{\Omega} \frac{1}{2}\left|\mathbf{u}^{n+1}  \right|^2 + C_0.
    \label{equ:alg_5}
  \end{equation}
  \begin{equation}
    \int_{\Omega} p^{n+1} = 0.
    \label{equ:alg_6}
  \end{equation}
\end{subequations}
%
$\Delta t$ is the time step
size in the above equations.
If $\chi$ denotes a generic variable, then in
the above equations
$
\frac{1}{\Delta t}(\gamma_0\chi^{n+1} - \hat{\chi})
$
is the approximation of
$\left.\frac{\partial\chi}{\partial t} \right|^{n+1}$
based on the $J$-th order 
backward differentiation formula (BDF), with
$\gamma_0$ and $\hat{\chi}$ defined specifically by
\begin{equation}
  \hat{\chi} = \left\{
  \begin{array}{ll}
    \chi^n, & J=1, \\
    2\chi^n - \frac{1}{2}\chi^{n-1}, & J=2;
  \end{array}
  \right.
  \qquad
  \gamma_0 = \left\{
  \begin{array}{ll}
    1, & J=1, \\
    3/2, & J=2.
  \end{array}
  \right.
  \label{equ:def_hat}
\end{equation}
%
$\bar{\mathbf{u}}^{n+1}$ denotes a $J$-th order explicit approximation
of $\mathbf{u}^{n+1}$, given by
  \begin{equation}
    \bar{\mathbf{u}}^{n+1} 
    = \left\{
    \begin{array}{ll}
      \mathbf{u}^n, & J=1, \\
      2\mathbf{u}^n - \mathbf{u}^{n-1}, & J=2.
    \end{array}
    \right.
    \label{equ:def_u_star}
  \end{equation}

%
%
%

Taking the $L^2$ inner product between $\mathbf{u}^{n+1}$
and equation \eqref{equ:alg_1} leads to
\begin{multline}
  \int_{\Omega}  \left[ 
    \frac{\gamma_0\mathbf{u}^{n+1}-\hat{\mathbf{u}}}{\Delta t}
    + \frac{R^{n+1}}{\sqrt{E^{n+1}}}\mathbf{N}(\bar{\mathbf{u}}^{n+1})
    \right] \cdot\mathbf{u}^{n+1}
  + \int_{\partial\Omega}(\mathbf{n}\cdot\mathbf{u}^{n+1})p^{n+1} \\
  -\nu\int_{\partial\Omega}(\mathbf{n}\cdot\nabla\mathbf{u}^{n+1})\cdot\mathbf{u}^{n+1}
  +\nu\int_{\Omega}\|\nabla\mathbf{u}^{n+1} \|^2
  = \int_{\Omega} \mathbf{f}^{n+1}\cdot\mathbf{u}^{n+1}
  \label{equ:nse_1_prod}
\end{multline}
where we have used integration by part, the divergence theorem,
and the equation \eqref{equ:alg_2}.
Sum up equations \eqref{equ:alg_3} and \eqref{equ:nse_1_prod},
and we have
\begin{multline}
  2R^{n+1}\frac{\gamma_0R^{n+1} - \hat{R}}{\Delta t}
  = -\nu\int_{\Omega}\|\nabla\mathbf{u}^{n+1} \|^2
  + \int_{\Omega} \mathbf{f}^{n+1}\cdot\mathbf{u}^{n+1} \\
  + \int_{\partial\Omega}\left(
  -p^{n+1}\mathbf{n}
  + \nu \mathbf{n}\cdot\nabla\mathbf{u}^{n+1}
  -\frac{1}{2}|\mathbf{w}^{n+1}|^2\mathbf{n}
  \right) \cdot \mathbf{w}^{n+1}
  \label{equ:eng_1}
\end{multline}
where we have used equation \eqref{equ:alg_4}.
Note the following relations
\begin{subequations}
  \begin{equation}
    2R^{n+1}(\gamma_0R^{n+1} - \hat{R})
    = 2R^{n+1}(R^{n+1}-R^n)
    = |R^{n+1}|^2 - |R^n|^2 + |R^{n+1}-R^n|^2,
    \quad \text{if} \ J=1;
    \label{equ:relation_1}
  \end{equation}
  \begin{equation}
    \begin{split}
    2R^{n+1}(\gamma_0R^{n+1} - \hat{R})
    =& 2R^{n+1}\left(\frac{3}{2}R^{n+1}-2R^n+\frac{1}{2}R^n\right) \\
    =& \frac{1}{2}\left(|R^{n+1}|^2 - |R^n|^2  \right)
    + \frac{1}{2}\left(\left|2R^{n+1}-R^n\right|^2 - \left|2R^n - R^{n-1}\right|^2  \right) \\
    &
    + \frac{1}{2}\left|R^{n+1}-2R^n + R^{n-1}\right|^2,
    \quad \text{if} \ J=2.
    \end{split}
  \end{equation}
\end{subequations}
Combining the above equations, we obtain the following stability
result about the scheme:
\begin{theorem}
  \label{thm:thm_1}
  
  In the absence of the external force $\mathbf{f}$ and with
  zero boundary velocity $\mathbf{w}$, the scheme represented by
  equations \eqref{equ:alg_1}--\eqref{equ:alg_6} satisfies the following
  property:
  \begin{equation}
    Q^{n+1} - Q^n = - \left|D^{n+1}\right|^2
    -\nu\Delta t \int_{\Omega}\|\nabla\mathbf{u}^{n+1} \|^2
    \label{equ:energy_balance}
  \end{equation}
  where
  \begin{equation}
    Q^{n} = \left\{
    \begin{array}{ll}
      |R^n|^2, & J=1, \\
      \frac{1}{2}\left(|R^n|^2 + |2R^n - R^{n-1}|^2  \right), & J=2;
    \end{array}
    \right.
    \ \
    D^{n+1} = \left\{
    \begin{array}{ll}
      R^{n+1} - R^n, & J=1, \\
      \frac{1}{\sqrt{2}}\left(R^{n+1} - 2R^n + R^{n-1}  \right), & J=2.
    \end{array}
    \right.
  \end{equation}
  
\end{theorem}


\subsection{Solution Algorithm and Implementation}

We next consider how to implement the algorithm
represented by the equations \eqref{equ:alg_1}--\eqref{equ:alg_6}.
Even though $R^{n+1}$ and $E^{n+1}$ are both implicit and
$E^{n+1}$ involves the integral of the unknown field function $\mathbf{u}^{n+1}$
over the domain, the scheme can be
implemented in an efficient way, thanks to
the fact that $R^{n+1}$ and $E^{n+1}$ are scalar numbers,
not field functions.
We employ $C^0$ spectral
elements~\cite{SherwinK1995,KarniadakisS2005,ZhengD2011,ChenSX2012} for spatial
discretizations in the current work.

Let
\begin{equation}
  S = \frac{R^{n+1}}{\sqrt{E^{n+1}}}, \quad
  E(S) = E^{n+1} = \int_{\Omega} \frac{1}{2}|\mathbf{u}^{n+1}|^2 +C_0.
  \label{equ:def_S}
\end{equation}
Then equation \eqref{equ:alg_1} can be written as
\begin{equation}
  \frac{\gamma_0}{\Delta t}\mathbf{u}^{n+1}
  + \nabla p^{n+1} - \nu\nabla^2\mathbf{u}^{n+1}
  = \mathbf{G}^{n+1} - S\mathbf{N}(\bar{\mathbf{u}}^{n+1})
  \label{equ:nse_trans_1}
\end{equation}
where
$\mathbf{G}^{n+1} = \mathbf{f}^{n+1} + \frac{\hat{\mathbf{u}}}{\Delta t}$.
In light of \eqref{equ:alg_2}, equation \eqref{equ:nse_trans_1}
can be transformed into
\begin{equation}
  \frac{\gamma_0}{\Delta t}\mathbf{u}^{n+1}
  + \nabla p^{n+1} 
  = \mathbf{G}^{n+1} - S\mathbf{N}(\bar{\mathbf{u}}^{n+1})
  - \nu\nabla\times\nabla\times\mathbf{u}^{n+1}.
  \label{equ:nse_trans_2}
\end{equation}

We would like to derive the weak forms for the pressure and
velocity in the spatially continuous space first. The discrete function
spaces for these variables will be specified later.
Let $q(\mathbf{x})$ denote an arbitrary test function
in the continuous space. 
Taking
the $L^2$ inner product between $\nabla q$ and equation \eqref{equ:nse_trans_2},
we get
\begin{equation}
  \int_{\Omega}\nabla p^{n+1} \cdot\nabla q
  = \int_{\Omega} \left[ \mathbf{G}^{n+1}
  - S\mathbf{N}(\bar{\mathbf{u}}^{n+1})\right]\cdot\nabla q
  -\nu\int_{\partial\Omega} \mathbf{n}\times\bm{\omega}^{n+1}\cdot\nabla q
  - \frac{\gamma_0}{\Delta t}\int_{\partial\Omega}\mathbf{n}\cdot\mathbf{w}^{n+1} q,
  \quad \forall q,
  \label{equ:p_weak_1}
\end{equation}
where $\bm{\omega}=\nabla\times\mathbf{u}$ is the vorticity,
  and we have used the integration by part, the divergence
theorem, equations \eqref{equ:alg_2} and \eqref{equ:alg_4}, and
the identity
$ 
  \int_{\Omega} \nabla\times\bm{\omega}\cdot\nabla q
  = \int_{\partial\Omega} \mathbf{n}\times\bm{\omega}\cdot\nabla q.
$ 
Equation \eqref{equ:p_weak_1} couples the pressure and the velocity
because of the vorticity $\bm{\omega}^{n+1}$ (tangent component)
on the boundary.
In order to simplify the implementation, we will make
the the following approximation,
$
\left.\mathbf{n}\times\bm{\omega}^{n+1}\right|_{\partial\Omega}
\approx \left.\mathbf{n}\times\bar{\bm{\omega}}^{n+1}\right|_{\partial\Omega}
$,
where
$
\bar{\bm{\omega}}^{n+1}
=\nabla\times \bar{\mathbf{u}}^{n+1}
$ is
the vorticity based on the explicitly approximated velocity
defined in \eqref{equ:def_u_star}.
This only slightly reduces the robustness in terms of stability,
but significantly simplifies the implementation and reduces
the computations. Employing this approximation,
we transform \eqref{equ:p_weak_1} into
\begin{equation}
  \int_{\Omega}\nabla p^{n+1} \cdot\nabla q
  = \int_{\Omega} \left[ \mathbf{G}^{n+1}
  - S\mathbf{N}(\bar{\mathbf{u}}^{n+1})\right]\cdot\nabla q
  -\nu\int_{\partial\Omega} \mathbf{n}\times\bar{\bm{\omega}}^{n+1}\cdot\nabla q
  - \frac{\gamma_0}{\Delta t}\int_{\partial\Omega}\mathbf{n}\cdot\mathbf{w}^{n+1} q,
  \quad \forall q.
  \label{equ:p_weakform}
\end{equation}
This equation, together with equation \eqref{equ:alg_6},
can be solved for $p^{n+1}$, provided that the unknown scalar value
$S=\frac{R^{n+1}}{\sqrt{E^{n+1}}}$ is given.


Exploiting the fact that $S$ is a scalar number (not a field function),
we solve equations \eqref{equ:p_weakform} and \eqref{equ:alg_6}
for $p^{n+1}$ as follows.
Define two field variables $p_1^{n+1}$ and $p_2^{n+1}$
as solutions to the following two problems: \\
\noindent\underline{For $p_1^{n+1}$:}
\begin{subequations}
  \begin{equation}
    \int_{\Omega}\nabla p_1^{n+1} \cdot\nabla q
  = \int_{\Omega} \mathbf{G}^{n+1}
  \cdot\nabla q
  -\nu\int_{\partial\Omega} \mathbf{n}\times\bar{\bm{\omega}}^{n+1}\cdot\nabla q
  - \frac{\gamma_0}{\Delta t}\int_{\partial\Omega}\mathbf{n}\cdot\mathbf{w}^{n+1} q,
  \quad \forall q.
  \label{equ:p1_1}
  \end{equation}
  \begin{equation}
    \int_{\Omega} p_1^{n+1} = 0.
    \label{equ:p1_2}
  \end{equation}
\end{subequations}
\noindent\underline{For $p_2^{n+1}$:}
\begin{subequations}
  \begin{equation}
    \int_{\Omega}\nabla p_2^{n+1} \cdot\nabla q
    = -\int_{\Omega} \mathbf{N}(\bar{\mathbf{u}}^{n+1})\cdot\nabla q,
    \quad \forall q.
  \label{equ:p2_1}
  \end{equation}
  \begin{equation}
    \int_{\Omega} p_2^{n+1} = 0.
    \label{equ:p2_2}
  \end{equation}
\end{subequations}
Then it is straightforward to verify that
the solution to equations \eqref{equ:p_weakform} and \eqref{equ:alg_6}
is given by
\begin{equation}
  p^{n+1} = p_1^{n+1} + Sp_2^{n+1}
  \label{equ:p_solution}
\end{equation}
where $S$ is to be determined later.

In light of \eqref{equ:p_solution}, we re-write
equation \eqref{equ:nse_trans_1} as
\begin{equation}
  \frac{\gamma_0}{\nu\Delta t}\mathbf{u}^{n+1} - \nabla^2\mathbf{u}^{n+1}
  =\frac{1}{\nu}(\mathbf{G}^{n+1} - \nabla p_1^{n+1})
  - \frac{S}{\nu}\left[\mathbf{N}(\bar{\mathbf{u}}^{n+1}) + \nabla p_2^{n+1}\right].
  \label{equ:nse_trans_3}
\end{equation}
Let $\varphi(\mathbf{x})$ denote an arbitrary test function
that vanishes on $\partial\Omega$, i.e.~$\left.\varphi\right|_{\partial\Omega}=0$.
Taking the $L^2$ inner product between $\varphi$ and
the equation \eqref{equ:nse_trans_3}, we obtain the weak form
about $\mathbf{u}^{n+1}$,
\begin{multline}
  \frac{\gamma_0}{\nu\Delta t}\int_{\Omega}\mathbf{u}^{n+1}\varphi
  + \int_{\Omega}\nabla\varphi\cdot\nabla\mathbf{u}^{n+1}
  = \frac{1}{\nu}\int_{\Omega}(\mathbf{G}^{n+1}-\nabla p_1^{n+1})\varphi \\
  - \frac{S}{\nu}\int_{\Omega}\left[
    \mathbf{N}(\bar{\mathbf{u}}^{n+1}) + \nabla p_2^{n+1}
    \right]\varphi,
  \quad \forall \varphi \ \text{with} \ \varphi|_{\partial\Omega}=0,
  \label{equ:u_weakform}
\end{multline}
where we have used integration by part, the divergence
theorem, and the fact that $\left.\varphi\right|_{\partial\Omega}=0$.
This equation, together with equation \eqref{equ:alg_4},
can be solved for $\mathbf{u}^{n+1}$,
provided that $S$ is known.

We again exploit the fact that $S$ is a scalar number, and
solve these equations for $\mathbf{u}^{n+1}$ as follows.
Define two field variables $\mathbf{u}_1^{n+1}$ and
$\mathbf{u}_2^{n+1}$ as solutions to the following two
problems: \\
\noindent\underline{For $\mathbf{u}_1^{n+1}$:}
\begin{subequations}
  \begin{equation}
    \frac{\gamma_0}{\nu\Delta t}\int_{\Omega}\mathbf{u}_1^{n+1}\varphi
  + \int_{\Omega}\nabla\varphi\cdot\nabla\mathbf{u}_1^{n+1}
  = \frac{1}{\nu}\int_{\Omega}(\mathbf{G}^{n+1}-\nabla p_1^{n+1})\varphi,
  \quad \forall \varphi \ \text{with} \ \varphi|_{\partial\Omega}=0.
  \label{equ:u1_1}
  \end{equation}
  \begin{equation}
    \mathbf{u}_1^{n+1} = \mathbf{w}^{n+1}, \quad \text{on} \ \partial\Omega.
    \label{equ:u1_2}
  \end{equation}
\end{subequations}
\noindent\underline{For $\mathbf{u}_2^{n+1}$:}
\begin{subequations}
  \begin{equation}
    \frac{\gamma_0}{\nu\Delta t}\int_{\Omega}\mathbf{u}_2^{n+1}\varphi
    + \int_{\Omega}\nabla\varphi\cdot\nabla\mathbf{u}_2^{n+1}
    = -\frac{1}{\nu}\int_{\Omega}\left[
      \mathbf{N}(\bar{\mathbf{u}}^{n+1}) + \nabla p_2^{n+1}
      \right] \varphi,
    \quad \forall \varphi \ \text{with} \ \varphi|_{\partial\Omega}=0.
    \label{equ:u2_1}
  \end{equation}
  \begin{equation}
    \mathbf{u}_2^{n+1} = 0, \quad \text{on} \ \partial\Omega.
    \label{equ:u2_2}
  \end{equation}
\end{subequations}
Then the solution $\mathbf{u}^{n+1}$ to equations \eqref{equ:u_weakform} and
\eqref{equ:alg_4} is given by
\begin{equation}
  \mathbf{u}^{n+1} = \mathbf{u}_1^{n+1} + S\mathbf{u}_2^{n+1},
  \label{equ:u_solution}
\end{equation}
where $S$ is to be determined.

We re-write equation \eqref{equ:alg_3} into
\begin{multline}
  \frac{2}{\Delta t}R^{n+1}(\gamma_0R^{n+1} - \hat{R})\frac{R^{n+1}}{\sqrt{E^{n+1}}}
  - \frac{\gamma_0}{\Delta t}\frac{R^{n+1}}{\sqrt{E^{n+1}}}\int_{\Omega}|\mathbf{u}^{n+1}|^2
  +\frac{1}{\Delta t}\frac{R^{n+1}}{\sqrt{E^{n+1}}} 
  \int_{\Omega}\hat{\mathbf{u}}\cdot\mathbf{u}^{n+1} \\
  - \left(\frac{R^{n+1}}{\sqrt{E^{n+1}}}\right)^2\int_{\Omega}\mathbf{N}(\bar{\mathbf{u}}^{n+1})\cdot\mathbf{u}^{n+1}
  + \frac{R^{n+1}}{\sqrt{E^{n+1}}} \int_{\partial\Omega}(\mathbf{n}\cdot\mathbf{u}^{n+1})
  \frac{1}{2}|\mathbf{u}^{n+1}|^2 = 0,
  \label{equ:R_equ_trans_1}
\end{multline}
where we have multiplied both sides of equation \eqref{equ:alg_3}
by $\frac{R^{n+1}}{\sqrt{E^{n+1}}}$.
We observe that this can improve the robustness of
the scheme when the time step size $\Delta t$
becomes large.
This is a scalar nonlinear equation, and it will be solved for
$S$.

In light of equation \eqref{equ:u_solution}, we have
\begin{equation}
  E^{n+1} = E(S) = C_0 + \int_{\Omega}\frac{1}{2}|\mathbf{u}^{n+1}|^2
  = A_0 + A_1S + A_2S^2,
  \label{equ:Es_equ}
\end{equation}
where
\begin{equation}
  \begin{split}
    A_0 = C_0 + \int_{\Omega}\frac{1}{2}|\mathbf{u}_1^{n+1}|^2, \quad
    A_1 = \int_{\Omega} \mathbf{u}_1^{n+1}\cdot\mathbf{u}_2^{n+1}, \quad
    A_2 = \int_{\Omega}\frac{1}{2}|\mathbf{u}_2^{n+1}|^2.
  \end{split}
  \label{equ:def_A012}
\end{equation}
In light of equations \eqref{equ:def_S}, \eqref{equ:alg_4}, \eqref{equ:p_solution}
and \eqref{equ:u_solution},
we can transform equation \eqref{equ:R_equ_trans_1} into
\begin{equation}
  F(S) = \frac{2\gamma_0}{\Delta t}S(S^2-1)E(S)
  -\frac{2\hat{R}}{\Delta t}S^2\sqrt{E(S)}
  + B_0S + B_1S^2 + B_2S^3
  = 0
  \label{equ:S_equ}
\end{equation}
where $\hat{R}$ is defined by \eqref{equ:def_hat} and
\begin{equation}
  \left\{
  \begin{split}
    &
    B_0 = \frac{2\gamma_0}{\Delta t}C_0
    + \frac{1}{\Delta t}\int_{\Omega}\hat{\mathbf{u}}\cdot\mathbf{u}_1^{n+1}
    + \int_{\partial\Omega}(\mathbf{n}\cdot\mathbf{w}^{n+1})\frac{1}{2}|\mathbf{w}^{n+1}|^2, \\
    &
    B_1 = \frac{1}{\Delta t}\int_{\Omega}\hat{\mathbf{u}}\cdot\mathbf{u}_2^{n+1}
    -\int_{\Omega} \mathbf{N}(\bar{\mathbf{u}}^{n+1})\cdot\mathbf{u}_1^{n+1}, \\
    &
    B_2 = -\int_{\Omega} \mathbf{N}(\bar{\mathbf{u}}^{n+1})\cdot\mathbf{u}_2^{n+1}.
  \end{split}
  \right.
  \label{equ:def_B012}
\end{equation}
Equation \eqref{equ:S_equ} is a nonlinear {\em scalar} equation about $S$.
It can be solved for $S$ using the Newton's method with an
initial guess $S=1$. The cost of this computation is very small
and is essentially negligible compared to the total cost within a time step,
which will be shown by the numerical experiments in Section \ref{sec:tests}.
With $S$ known, $E^{n+1}$ can be computed based on equation \eqref{equ:Es_equ},
and $R^{n+1}$ can be computed using equation \eqref{equ:def_S}.
The velocity $\mathbf{u}^{n+1}$ and pressure $p^{n+1}$
are then given by equations \eqref{equ:u_solution}
and \eqref{equ:p_solution}.


Let us now consider the spatial discretization of the equations
\eqref{equ:p1_1}--\eqref{equ:p2_2} and \eqref{equ:u1_1}--\eqref{equ:u2_2}.
We discretize the domain $\Omega$ with a mesh consisting
of $N_{el}$ conforming spectral elements.
Let the positive integer $K$ denote the element order,
which represents a measure of the highest polynomial degree
in the polynomial expansions of the field variables
within an element. Let $\Omega_h$ denote the
discretized domain, $\partial\Omega_h$ denote the boundary of $\Omega_h$,
and $\Omega_h^e$ ($1\leqslant e\leqslant N_{el}$)
denote the element $e$.
Define two function spaces
\begin{equation}
  \left\{
  \begin{split}
    &
  X = \{\
  v \in H^1(\Omega_h) \ :\
  v \ \text{is a polynomial characterized by} \ K\ \text{on}\ \Omega_h^e,
  \ \text{for} \ 1\leqslant e\leqslant N_{el}
  \ \}, \\
  &
  X_0 = \{\
  v \in X \ :\
  \left. v\right|_{\partial\Omega_h} = 0
  \ \}.
  \end{split}
  \right.
\end{equation}

In the following let $d$ ($d=2$ or $3$) denote the spatial dimension,
and the subscript in $(\cdot)_h$ denote
the discretized version of the variable $(\cdot)$.
The fully discretized equations corresponding to
\eqref{equ:p1_1}--\eqref{equ:p2_2} are: \\
\noindent\underline{For $p_{1h}^{n+1}$:} \ \
find $p_{1h}^{n+1}\in X$ such that
\begin{subequations}
  \begin{equation}
    \int_{\Omega_h}\nabla p_{1h}^{n+1} \cdot\nabla q_h
  = \int_{\Omega_h} \mathbf{G}_h^{n+1}
  \cdot\nabla q_h
  -\nu\int_{\partial\Omega_h} \mathbf{n}_h\times\bar{\bm{\omega}}_h^{n+1}\cdot\nabla q_h
  - \frac{\gamma_0}{\Delta t}\int_{\partial\Omega_h}\mathbf{n}_h\cdot\mathbf{w}_h^{n+1} q_h,
  \ \forall q_h\in X.
  \label{equ:p1_1_disc}
  \end{equation}
  \begin{equation}
    \int_{\Omega_h} p_{1h}^{n+1} = 0.
    \label{equ:p1_2_disc}
  \end{equation}
\end{subequations}
\noindent\underline{For $p_{2h}^{n+1}$:} \ \
find $p_{2h}^{n+1}\in X$ such that
\begin{subequations}
  \begin{equation}
    \int_{\Omega_h}\nabla p_{2h}^{n+1} \cdot\nabla q_h
    = -\int_{\Omega_h} \mathbf{N}(\bar{\mathbf{u}}_h^{n+1})\cdot\nabla q_h,
    \quad \forall q_h\in X.
  \label{equ:p2_1_disc}
  \end{equation}
  \begin{equation}
    \int_{\Omega_h} p_{2h}^{n+1} = 0.
    \label{equ:p2_2_disc}
  \end{equation}
\end{subequations}
The fully discretized equations corresponding to
\eqref{equ:u1_1}--\eqref{equ:u2_2} are: \\
\noindent\underline{For $\mathbf{u}_{1h}^{n+1}$:} \ \
find $\mathbf{u}_{1h}^{n+1}\in [X]^d$ such that
\begin{subequations}
  \begin{equation}
    \frac{\gamma_0}{\nu\Delta t}\int_{\Omega_h}\mathbf{u}_{1h}^{n+1}\varphi_h
  + \int_{\Omega_h}\nabla\varphi_h\cdot\nabla\mathbf{u}_{1h}^{n+1}
  = \frac{1}{\nu}\int_{\Omega_h}\left(\mathbf{G}_h^{n+1}-\nabla p_{1h}^{n+1}\right)\varphi_h,
  \quad \forall \varphi_h \in X_0.
  \label{equ:u1_1_disc}
  \end{equation}
  \begin{equation}
    \mathbf{u}_{1h}^{n+1} = \mathbf{w}_h^{n+1}, \quad \text{on} \ \partial\Omega_h.
    \label{equ:u1_2_disc}
  \end{equation}
\end{subequations}
\noindent\underline{For $\mathbf{u}_{2h}^{n+1}$:} \ \
find $\mathbf{u}_{2h}^{n+1}\in [X]^d$ such that 
\begin{subequations}
  \begin{equation}
    \frac{\gamma_0}{\nu\Delta t}\int_{\Omega_h}\mathbf{u}_{2h}^{n+1}\varphi_h
    + \int_{\Omega_h}\nabla\varphi_h\cdot\nabla\mathbf{u}_{2h}^{n+1}
    = -\frac{1}{\nu}\int_{\Omega_h}\left[
      \mathbf{N}(\bar{\mathbf{u}}_h^{n+1}) + \nabla p_{2h}^{n+1}
      \right]\varphi_h,
    \quad \forall \varphi_h \in X_0.
    \label{equ:u2_1_disc}
  \end{equation}
  \begin{equation}
    \mathbf{u}_{2h}^{n+1} = 0, \quad \text{on} \ \partial\Omega_h.
    \label{equ:u2_2_disc}
  \end{equation}
\end{subequations}


Combining the above discussions, we arrive at the final solution algorithm.
It involves the following steps: 
\begin{enumerate}[(i)]

\item
  Solve equations \eqref{equ:p1_1_disc}--\eqref{equ:p1_2_disc} for $p_1^{n+1}$; \\
  Solve equations \eqref{equ:u1_1_disc}--\eqref{equ:u1_2_disc} for $\mathbf{u}_1^{n+1}$.


\item
  Solve equations \eqref{equ:p2_1_disc}--\eqref{equ:p2_2_disc} for $p_2^{n+1}$; \\
  Solve equations \eqref{equ:u2_1_disc}--\eqref{equ:u2_2_disc} for $\mathbf{u}_2^{n+1}$.

\item
  Compute the coefficients $A_0$, $A_1$ and $A_2$ based on \eqref{equ:def_A012}; \\
  Compute the coefficients $B_0$, $B_1$ and $B_2$ based on \eqref{equ:def_B012}.

\item
  Solve equation \eqref{equ:S_equ} for $S$ using the Newton's method
  with an initial guess $S=1$.

\item
  Compute $\mathbf{u}^{n+1}$ from equation \eqref{equ:u_solution}; \\
  Compute $p^{n+1}$ from equation \eqref{equ:p_solution}; \\
  Compute $E^{n+1}$ from equation \eqref{equ:Es_equ}; \\
  Compute $R^{n+1}$ by $R^{n+1} = S\sqrt{E^{n+1}}$.
  
\end{enumerate}
It is noted that the linear algebraic systems for
$p_1^{n+1}$, $p_2^{n+1}$, $\mathbf{u}_1^{n+1}$, and $\mathbf{u}_2^{n+1}$
all involve constant and time-independent coefficient matrices,
which can be pre-computed during pre-processing.

%

\section{Representative Numerical Examples}
\label{sec:tests}

We next use several numerical examples in two dimensions 
to test the performance and the accuracy
of the algorithm presented in the previous section.
The spatial and temporal convergence rates of the scheme
will first be demonstrated using a contrived analytic
solution to the incompressible Navier-Stokes equations.
Then we will employ the method
to study a steady-flow problem (Kovasznay flow)
and the flow past a circular cylinder in a periodic channel
for a range of Reynolds numbers.

\subsection{Convergence Rates}

In this subsection we employ a manufactured analytic solution
to the incompressible Navier-Stokes equations to investigate 
the spatial and temporal convergence rates of the algorithm
developed herein.

\begin{figure}
  \centerline{
    \includegraphics[width=3in]{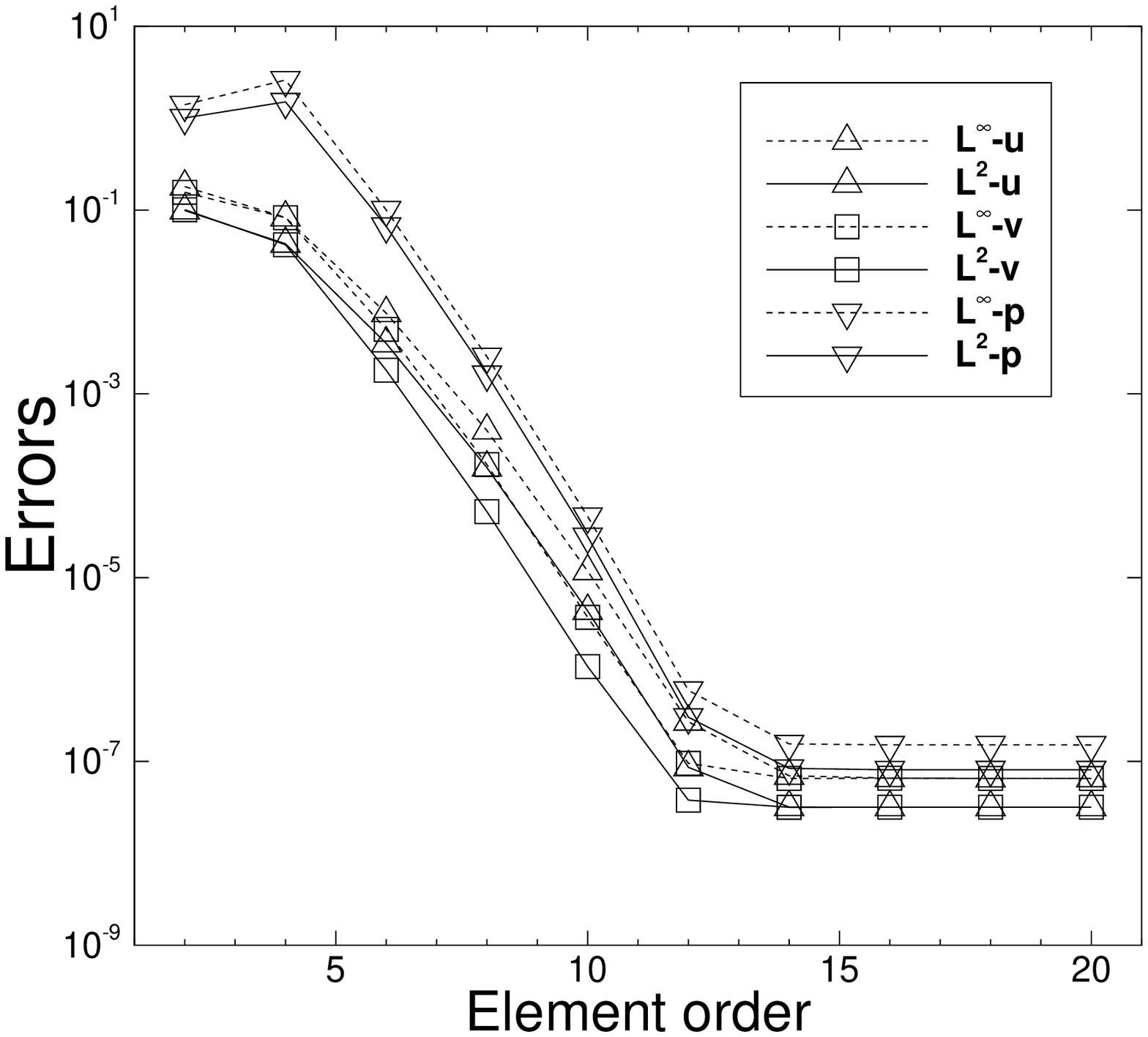}(a)
    \includegraphics[width=3in]{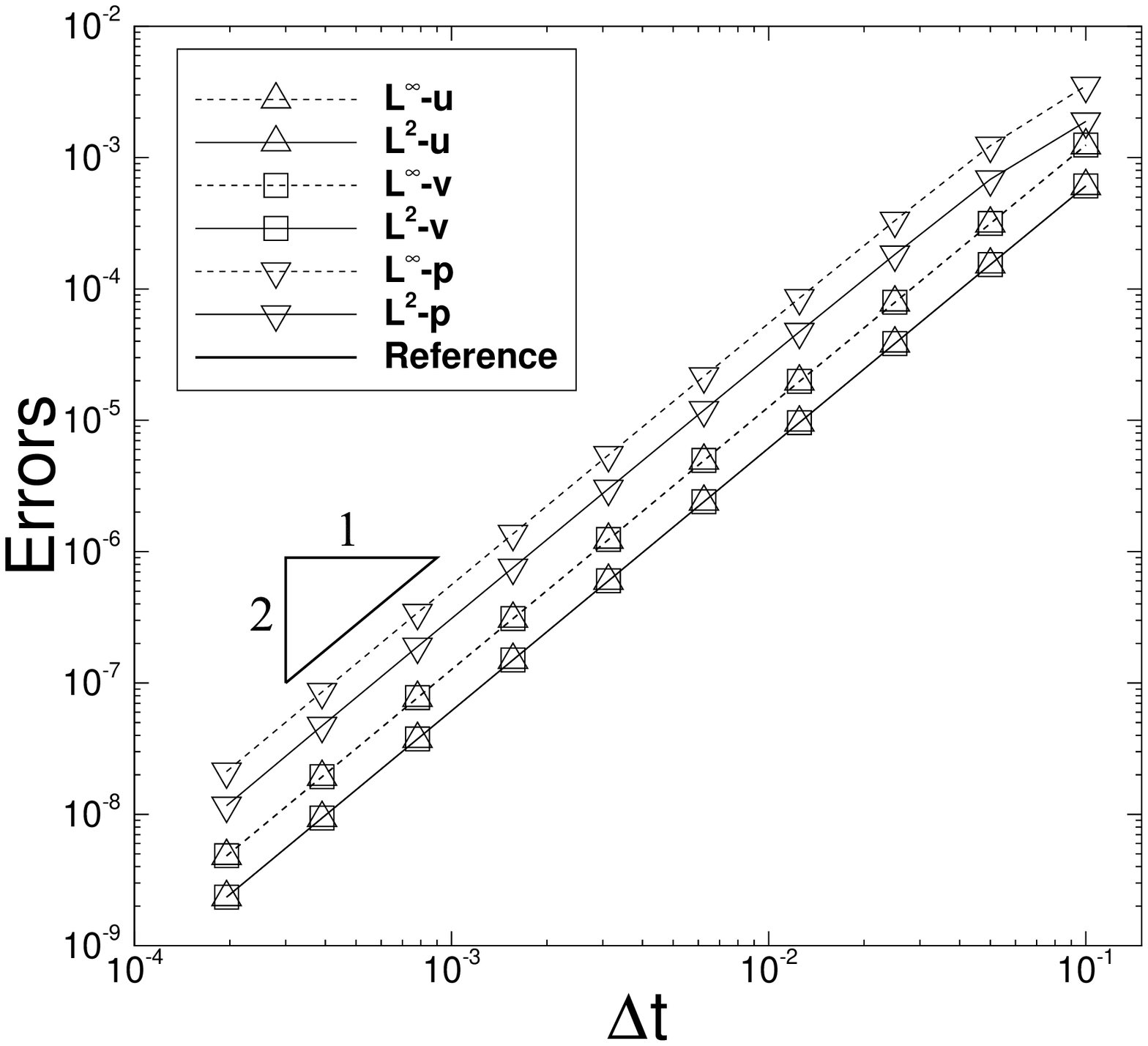}(a)
  }
  \caption{
    Convergence tests:
    (a) Numerical errors vs.~element order (fixed $\Delta t=0.001$ and $t_f=0.1$).
    (b) Numerical errors vs.~$\Delta t$ (fixed element order $16$ and $t_f=0.2$).
  }
  \label{fig:conv}
\end{figure}

Consider a rectangular domain $0\leqslant x\leqslant 2$ and $-1\leqslant y\leqslant 1$,
and the following solution to the Navier-Stokes equations 
\eqref{equ:nse}--\eqref{equ:continuity} and \eqref{equ:p_condition} on this domain:
\begin{equation}
\left\{
\begin{split}
&
u = 2\cos(\pi y)\sin(\pi x)\sin t, \\
&
v = -2\sin(\pi y)\cos(\pi x)\sin t, \\
&
p = 2\sin(\pi y)\sin(\pi x)\cos t,
\end{split}
\right.
\label{equ:anal_soln}
\end{equation}
where $u$ and $v$ are the two components of 
the velocity $\mathbf{u}$.
In equation \eqref{equ:nse} the external body force $\mathbf{f}$
is chosen such that the expressions in \eqref{equ:anal_soln}
satisfy this equation. 

We simulate this problem 
using the algorithm from
Section \ref{sec:method}. The domain is discretized
with two uniform elements along the $x$ direction.
On the domain boundary we impose the Dirichlet boundary
condition \eqref{equ:bc_vel}, in which the boundary velocity
 $\mathbf{w}$ is chosen based on the analytic
expressions in \eqref{equ:anal_soln}.
The initial velocity $\mathbf{u}_{in}$ is chosen according to
the analytic expressions in \eqref{equ:anal_soln} by 
setting $t=0$.

We integrate the Navier-Stokes equations in time
from $t=0$ to $t=t_f$ ($t_f$ to be specified next).
Then we compare the numerical solutions of different
flow variables at $t=t_f$ with the analytic expressions
from \eqref{equ:anal_soln}, and the errors in different
norms are computed and recorded.
The element order and the time step size $\Delta t$
are varied systematically in the simulations in order to study
their effects on the numerical errors.
We employ a non-dimensional
viscosity $\nu = 0.01$ for this problem, and 
the constant $C_0$ in equation \eqref{equ:def_energy}
is fixed at $C_0=1.0$ in the simulations.

In the spatial convergence tests, we employ
a fixed $t_f=0.1$ and $\Delta t=0.001$, and vary
the element order systematically between $2$ and $20$.
The errors at $t=t_f$ between the numerical solution and
the analytic solution in $L^{\infty}$ and $L^2$
norms have been computed corresponding to
all these element orders.
Figure \ref{fig:conv}(a) shows these numerical errors
as a function of the element order for this group
of tests. For element orders $12$ and below we observe
an exponential decrease in the numerical errors
as the element order increases.
As the element order increases beyond $14$, we observe
a saturation in the numerical errors due to
the temporal truncation error. The error curves
for different flow variables level off
approximately at a level around $10^{-7}$.
These results are indicative of an exponential
convergence rate in space for the current method.

In the temporal convergence tests, we fix the integration time at
$t_f=0.2$ and the element order at a large value $16$,
and then vary the time step size systematically
between $\Delta t=0.1$ and
$\Delta t=1.953125\times 10^{-4}$.
Figure \ref{fig:conv}(b) shows the $L^{\infty}$ and $L^2$ errors
of the velocity and pressure as a function of $\Delta t$,
plotted in logarithmic scales for both axes.
A second-order convergence rate in time is clearly observed
with the method developed herein.


\subsection{Kovasznay Flow}
\label{sec:kovas}

In this subsection we test the proposed method using a steady-state
problem, the Kovasznay flow~\cite{Kovasznay1948}, for which the
exact solution of the flow field is available.


\begin{figure}
  \centerline{
    \includegraphics[width=3in]{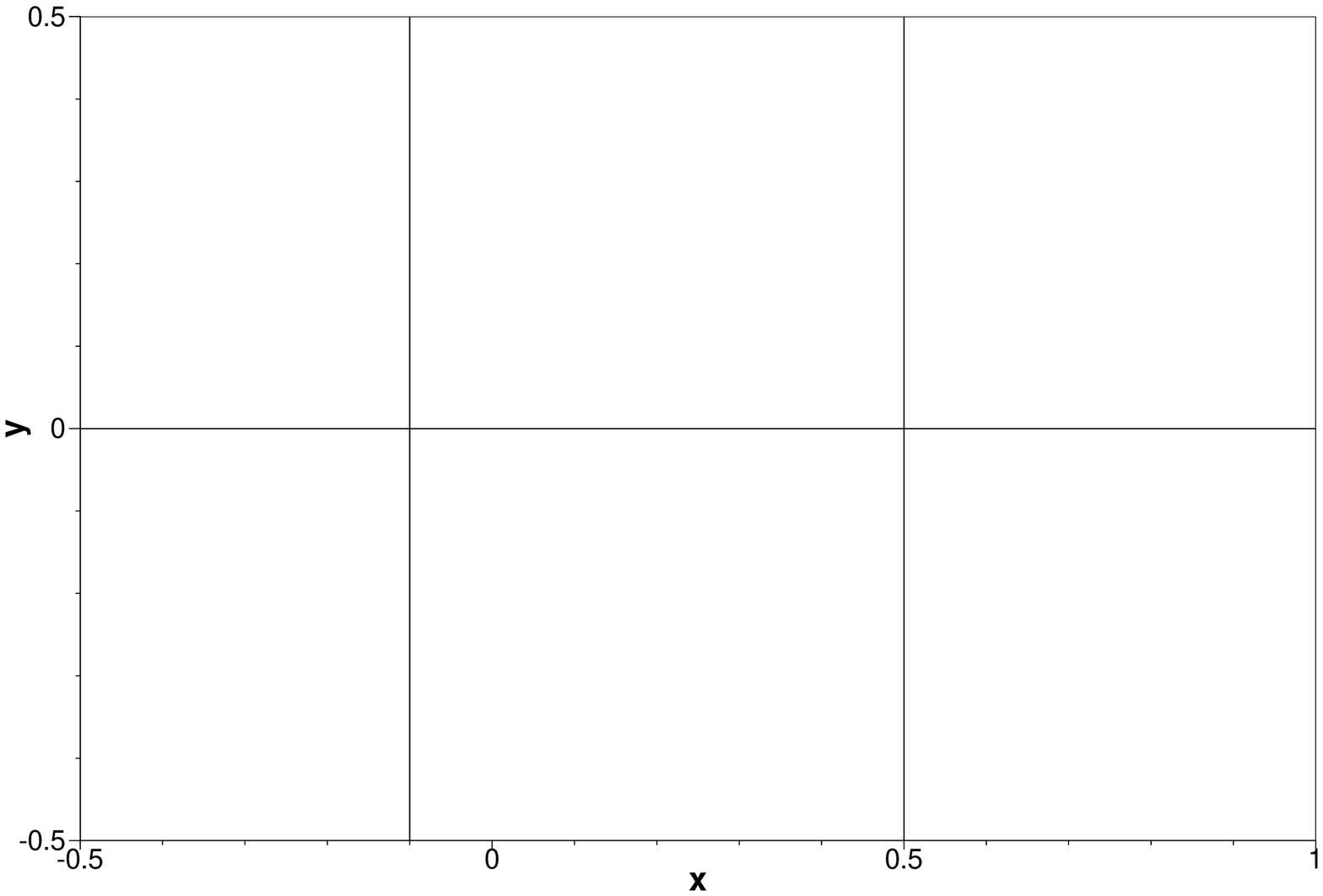}(a)
    \includegraphics[width=3in]{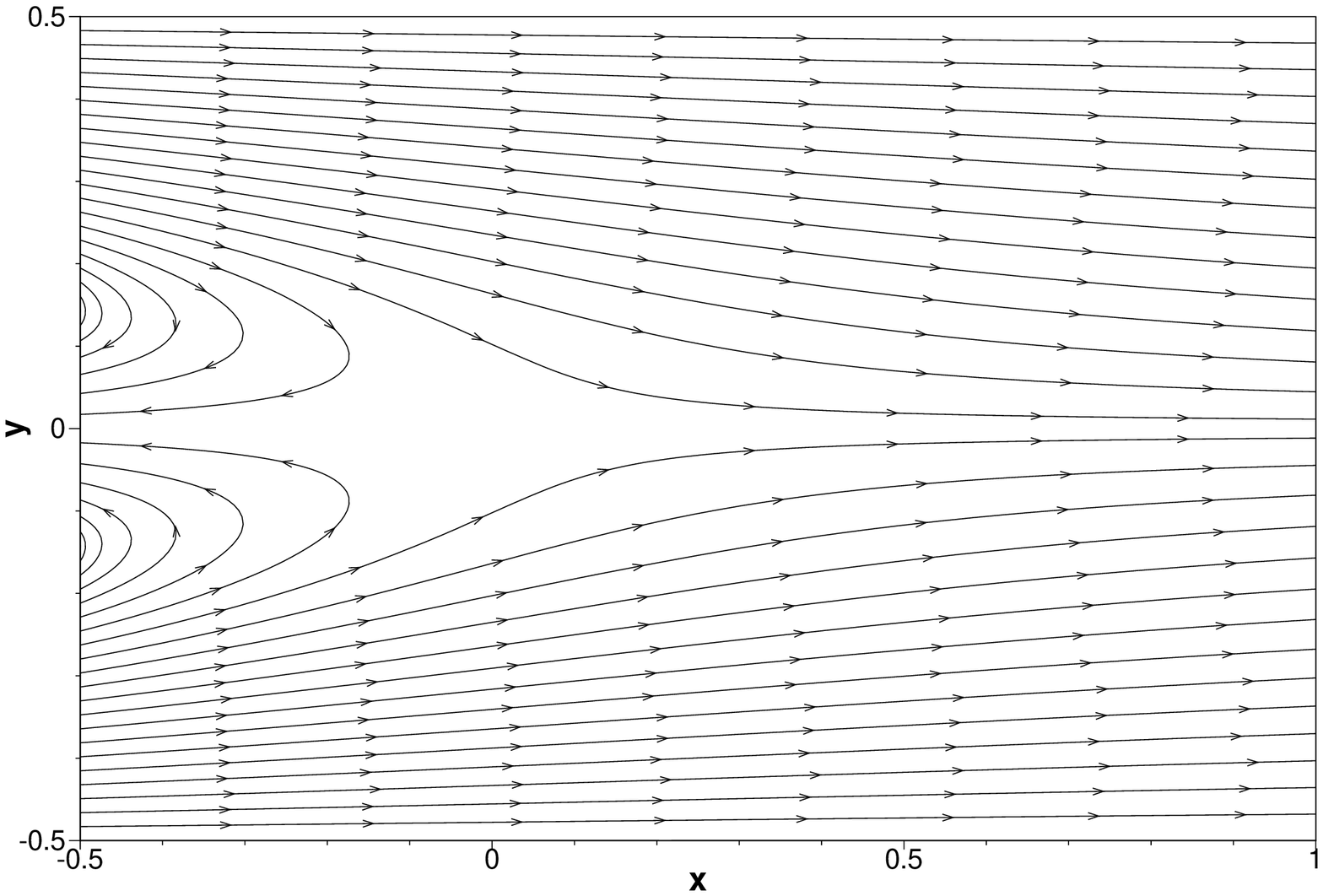}(b)
  }
  \caption{
    Kovasznay flow: (a) Flow domain and mesh of $6$ quadrilateral elements;
    (b) Streamlines.
  }
  \label{fig:kovas_config}
\end{figure}

Specifically, we consider the domain $0\leqslant x\leqslant 1$
and $-0.5\leqslant y\leqslant 0.5$, as depicted in
Figure \ref{fig:kovas_config}(a). The Kovasznay flow solution to
the Navier-Stokes equations \eqref{equ:nse}--\eqref{equ:continuity}
(with $\mathbf{f}=0$)
is given by the following expressions for the
velocity $\mathbf{u}=(u,v)$ and pressure:
\begin{equation}
  \left\{
  \begin{split}
    &
    u = 1-e^{\lambda x}\cos(2\pi y), \\
    &
    v = \frac{\lambda}{2\pi} e^{\lambda x}\sin(2\pi y), \\
    &
    p = \frac{1}{2}(1-e^{2\lambda x}),
  \end{split}
  \right.
  \label{equ:kovas_soln}
\end{equation}
where the constant
$
\lambda = \frac{1}{2\nu} - \sqrt{\frac{1}{4\nu^2} + 4\pi^2}.
$
The flow pattern for this solution is illustrated by the streamlines
shown in Figure \ref{fig:kovas_config}(b), which is similar to that
behind an obstacle.
We employ a non-dimensional viscosity $\nu = 0.025$ for this problem.


To simulate the problem,
we discretize the domain using $6$ quadrilateral elements
as shown in Figure \ref{fig:kovas_config}(a).
The element orders and the time step sizes are varied in the tests
and will be specified
subsequently. In the Navier-Stokes equation \eqref{equ:nse}
the external body force is set to $\mathbf{f}=0$.
On the four boundaries Dirichlet type 
conditions are imposed for the velocity according to
the expression given in \eqref{equ:kovas_soln}.
We set a zero initial velocity, $\mathbf{u}_{in}=0$,
in the initial condition \eqref{equ:ic_vel}.
The simulations have been performed for a sufficiently long
time until the steady state is reached.
Then the errors of the numerical solutions against
the exact solution as given by \eqref{equ:kovas_soln}
are computed, as well as the norms of the flow
variables.


\begin{figure}
  \centerline{
    \includegraphics[width=3.0in]{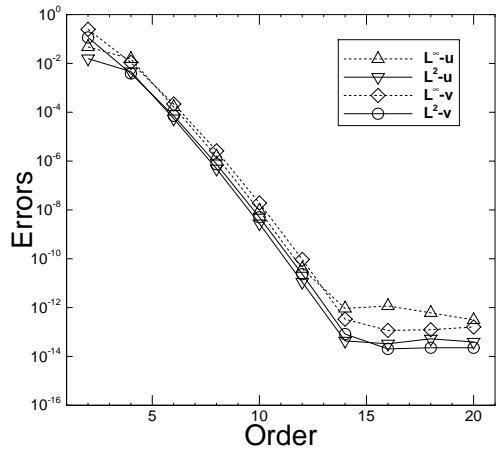}
  }
  \caption{ Kovasznay flow:
    $L^{\infty}$ and $L^2$ errors of the steady-state velocity
    as a function of the element order. Results are obtained with
    $\Delta t=0.005$ and $C_0=0.01$.
  }
  \label{fig:kovas_conv}
\end{figure}

Figure \ref{fig:kovas_conv} shows the $L^{\infty}$ and $L^2$
errors of the steady-state velocity from the simulations 
as a function of the element order. In this group of tests
the time step size is fixed at $\Delta t=0.005$, and
the constant $C_0$ in equation \eqref{equ:def_energy} is
$C_0=0.01$, while the element order has been varied
systematically between $2$ and $20$.
The errors of the steady-state velocity decrease
exponentially with increasing element order, until they saturate
at a level about $10^{-13}$ as the element order increases to $14$ and beyond.


\begin{figure}
  \centerline{
    \includegraphics[width=3in]{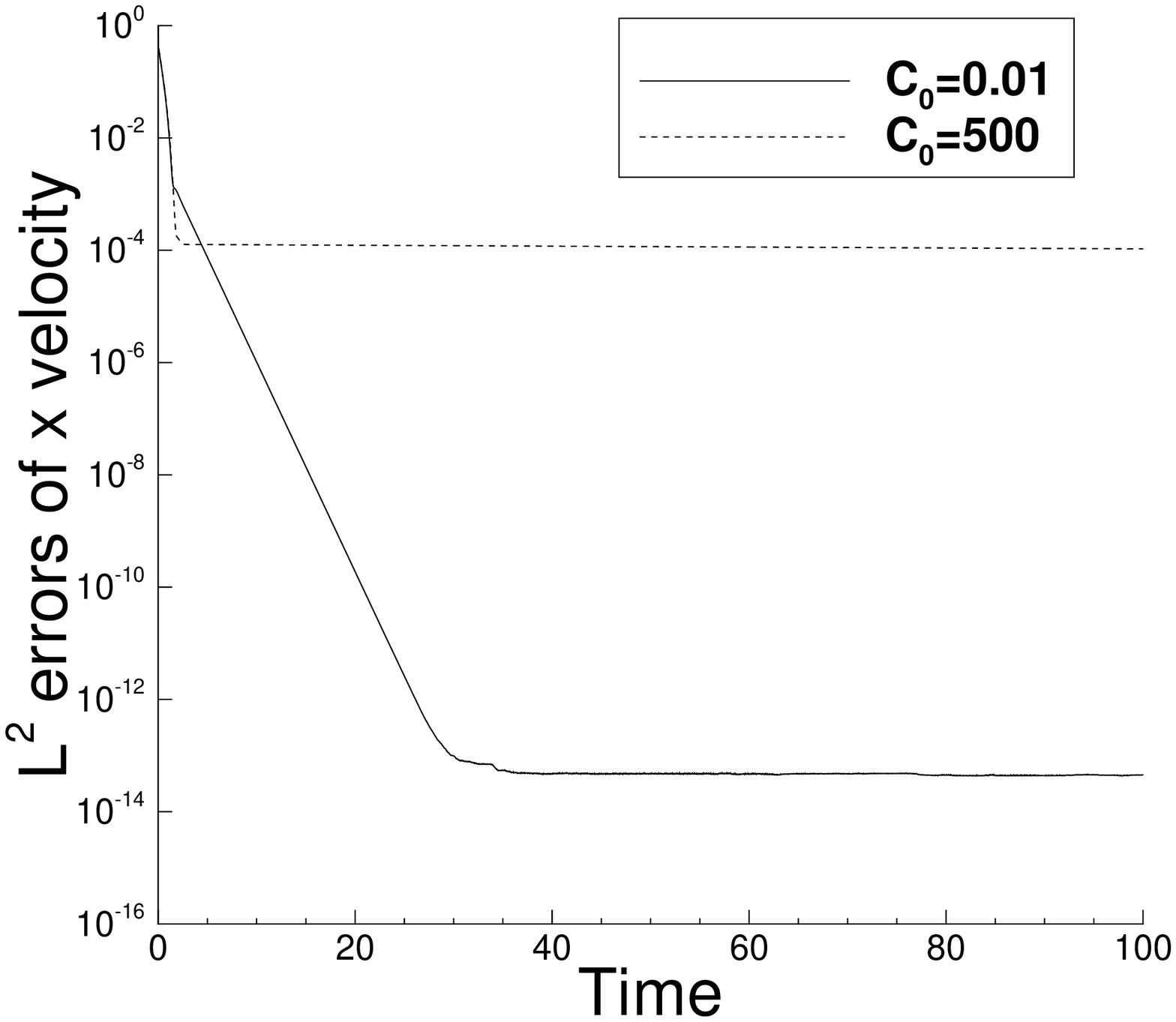}(a)
    \includegraphics[width=3in]{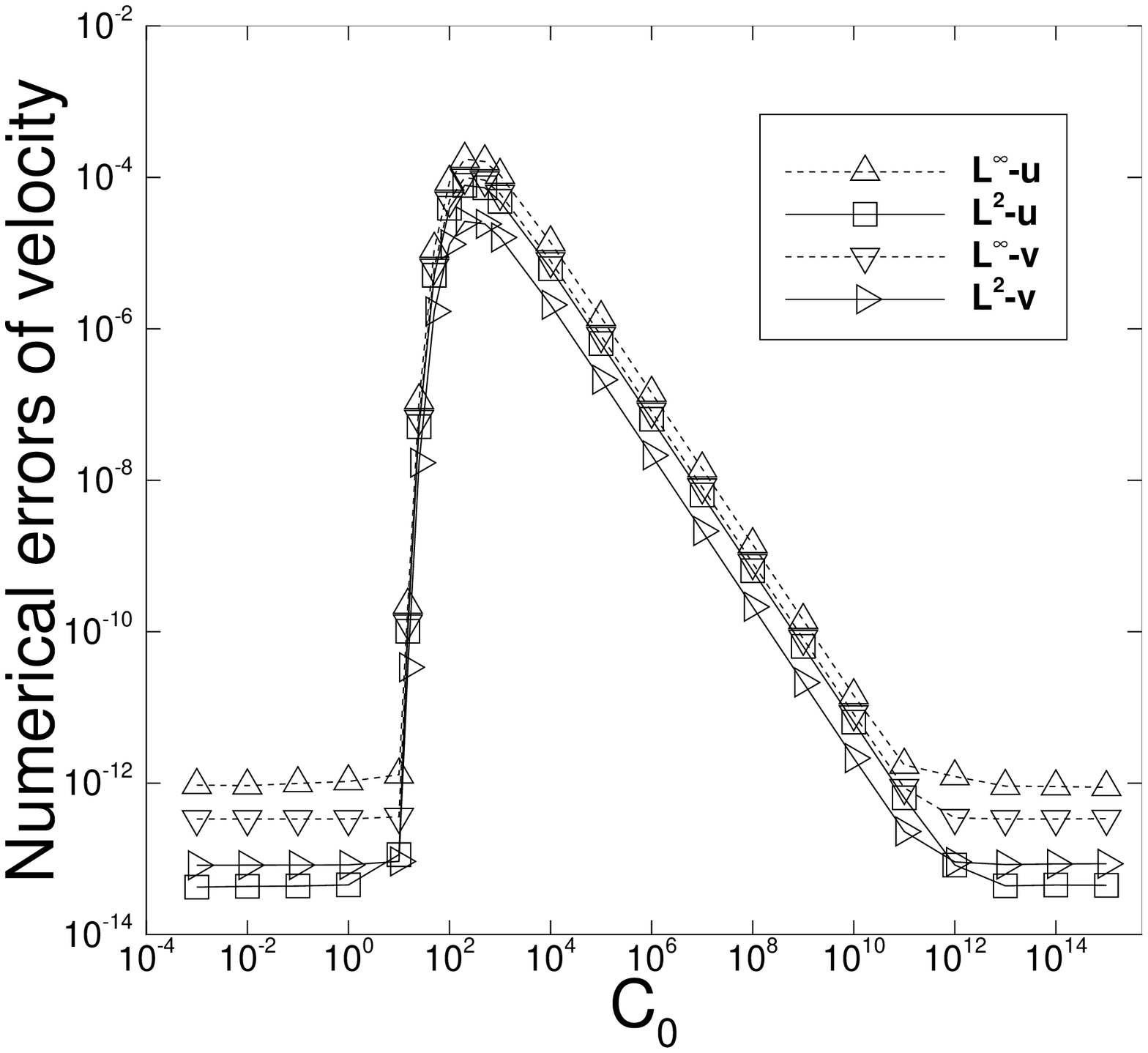}(b)
  }
  \caption{
    Effect of $C_0$ value on the simulation errors of Kovasznay flow:
    (a) Time histories of $L^2$ errors of the $x$ velocity
    obtained with $C_0=0.01$ and $C_0=500$.
    (b) Numerical errors of the steady-state velocities
    as a function of $C_0$.
    Results are obtained using
    $\Delta t=0.005$ and an element order $14$.
  }
  \label{fig:kovas_c0}
\end{figure}

We observe that the value for the constant $C_0$ in \eqref{equ:def_energy}
has a marked influence on the accuracy of the solution to
the steady-state velocity for this problem.
This is demonstrated by the results
in Figure \ref{fig:kovas_c0}. Figure \ref{fig:kovas_c0}(a)
shows the $L^2$ errors of the $x$ velocity
component as a function of time  as the simulation proceeds.
These results are for 
two constant values, $C_0=0.01$ and $C_0=500$.
The time step size is $\Delta t=0.005$ and the element order
is $14$ in these simulations.
An initial reduction in the numerical error is evident from
both history curves. But after this initial stage the
error corresponding to $C_0=500$ levels off at a value around $10^{-4}$,
while for $C_0=0.01$ the error levels off at a value between
$10^{-14}$ and $10^{-13}$. 
In light of this difference in the error  levels,
we have studied the $C_0$ effect more systematically.
We vary $C_0$ between $0.001$ and $10^{15}$, and
for each $C_0$ value the errors of
the steady-state velocity is computed and recorded.
In Figure \ref{fig:kovas_c0}(b) we show the errors of the steady-state
velocity as a function of $C_0$.
Fixed values of $\Delta t=0.005$ and an
element order $14$ have been employed in these tests.
%
As $C_0$ increases from $10^{-3}$ to about $10$, the numerical errors
remain essentially the same (around $10^{-13}$).
Then as $C_0$ increases from $10$ to about $500$ there is
a sharp increase in the numerical errors,
and the errors peak at a $C_0$ value
between $200$ and $500$ (reaching a level of about $10^{-4}$).
Beyond $C_0=500$, the numerical errors decrease with increasing
$C_0$, and the errors are approximately inverse proportional to
$C_0$ between $C_0=1000$ and $C_0=10^{11}$.
As $C_0$ increases to $10^{12}$ and beyond, the numerical errors
remain essentially the same, at a level around $10^{-13}$.
These results suggest that using a small $C_0$ or a very large
$C_0$ in the algorithm seems more favorable in terms of the accuracy.
We employ a constant value $C_0=0.01$ for all the results reported
subsequently in this subsection.
%


\begin{table}
  \begin{center}
    \begin{tabular}{lll}
      \hline
      $\Delta t$ & Element order $10$ & Element order $16$ \\
      $0.001$ & $2.61e-9$ & $9.67e-14$ \\
      $0.002$ & $2.69e-9$ & $9.50e-14$ \\
      $0.003$ & $2.72e-9$ & $6.98e-14$ \\
      $0.004$ & $2.74e-9$ & $3.87e-14$ \\
      $0.005$ & $2.76e-9$ & $3.15e-14$ \\
      $0.006$ & $2.77e-9$ & $6.81e-3$ \\
      $0.007$ & $2.78e-9$ & $1.91e-2$ \\
      $0.008$ & $2.79e-9$ & $2.58e-2$ \\
      $0.009$ & $2.80e-9$ & $3.64e-2$ \\
      $0.01$ & $1.36e-2$ & $3.83e-2$ \\
      $0.03$ & $1.27e-1$ & $1.17e-1$ \\
      $0.05$ & $1.54e-1$ & $1.41e-1$ \\
      $0.1$ & $1.78e-1$ & $1.85e-1$ \\
      $0.5$ & $2.74e-1$ & $2.61e-1$ \\
      $1.0$ & $2.95e-1$ & $2.95e-1$ \\
      \hline
    \end{tabular}
  \end{center}
  \caption{
    Kovasznay flow: $L^2$ errors of $x$ component of steady-state velocity versus
    $\Delta t$, computed with fixed $C_0=0.01$ and element orders $10$ and $16$.
  }
  \label{tab:kovas_dt}
\end{table}


\begin{figure}
  \centerline{
    \includegraphics[width=3in]{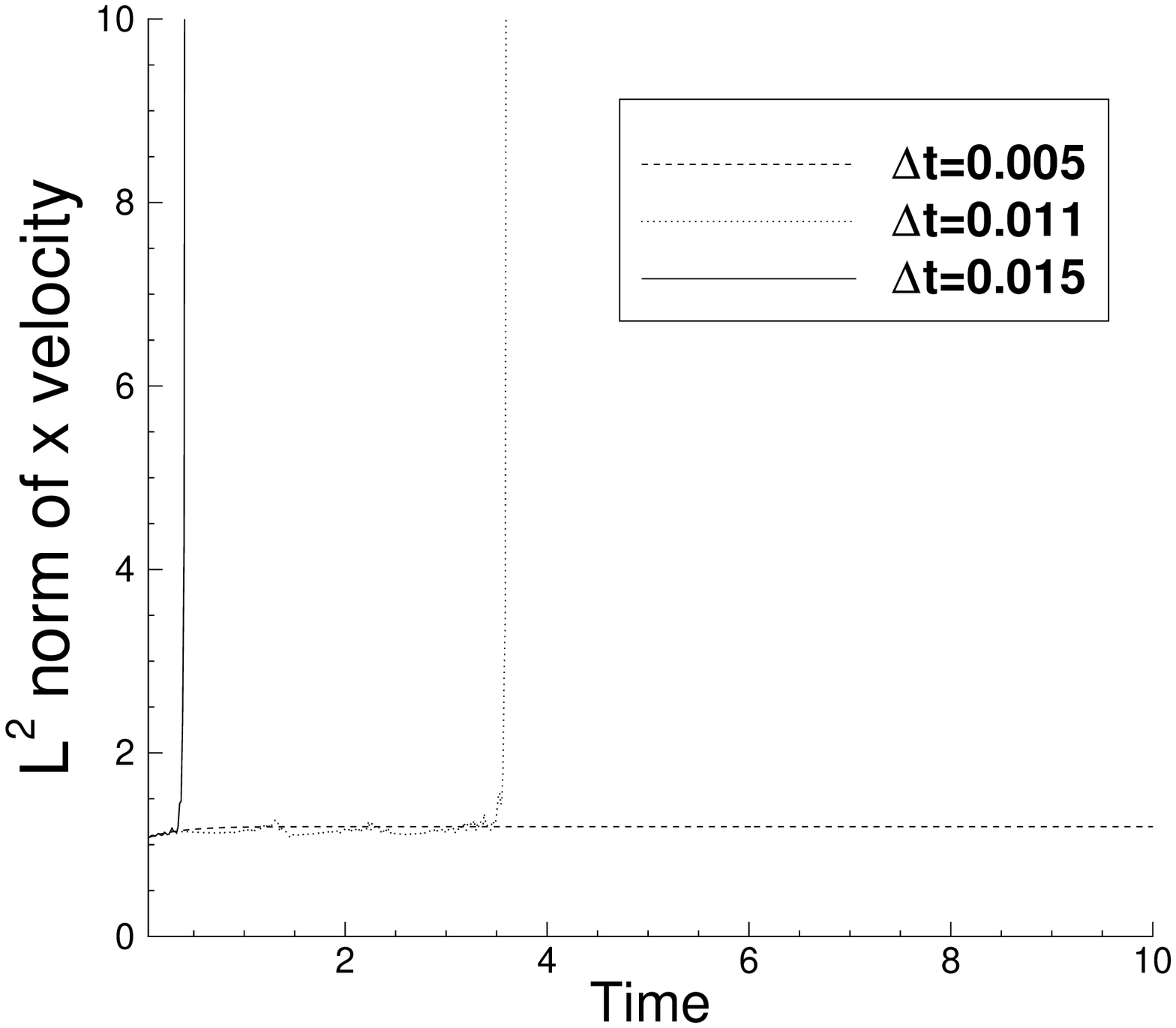}(a)
    \includegraphics[width=3in]{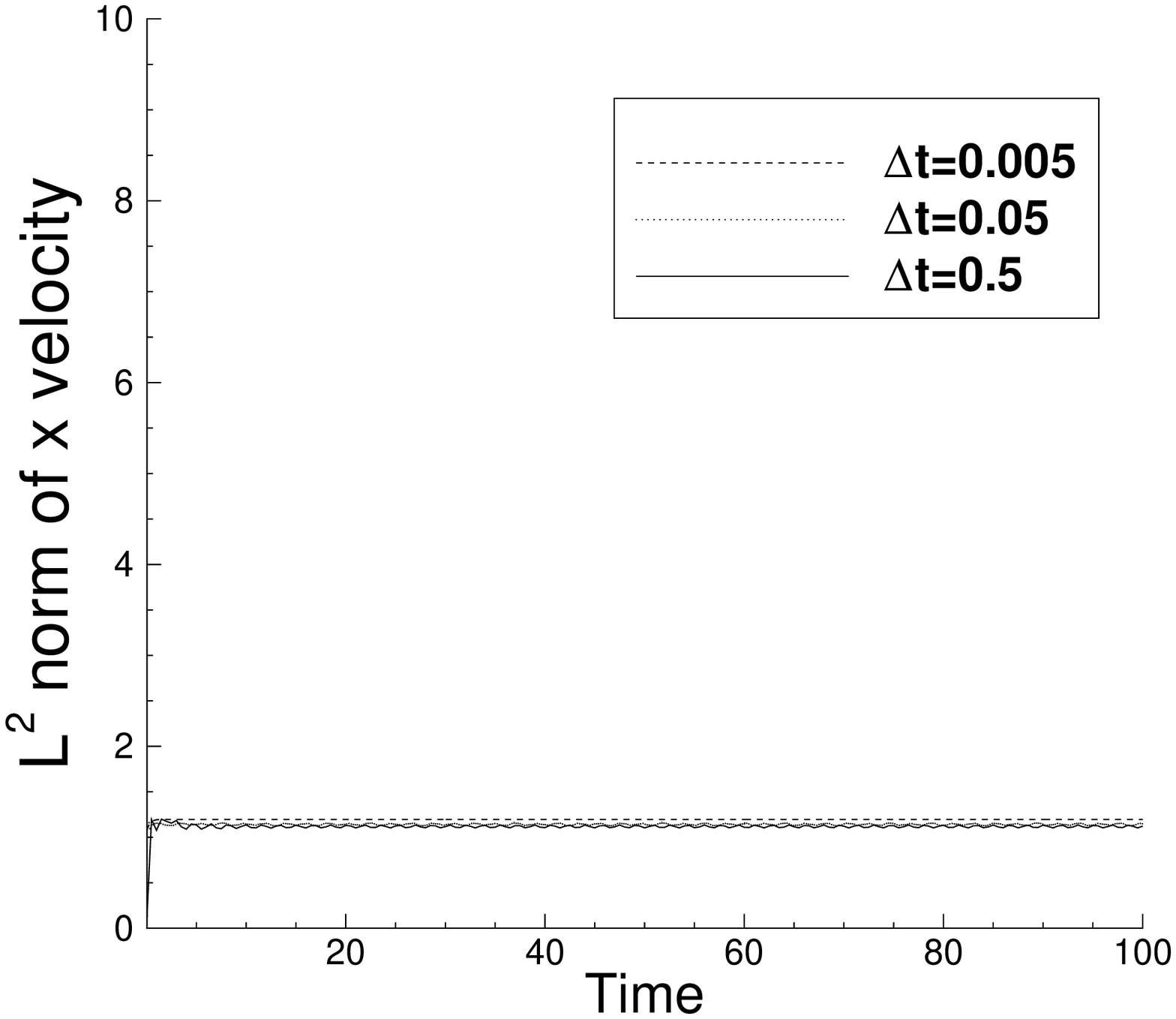}(b)
  }
  \caption{
    Kovasznay flow:
    Time histories of the $L^2$ norm of the $x$ velocity obtained using
    several $\Delta t$ values using
    (a) the semi-implicit scheme of \cite{Dong2015clesobc}, and
    (b) the current scheme.
  }
  \label{fig:kovas_semiimp}
\end{figure}


The energy stability property of the current scheme (see Theorem \ref{thm:thm_1})
is conducive to the stability of computations. We observe that
this scheme allows the use of fairly large or large time step sizes ($\Delta t$)
in the simulations, at least for steady-state problems.
In Table \ref{tab:kovas_dt} we have listed the $L^2$ errors of
the $x$ component of the steady-state velocity computed with
different $\Delta t$ values, ranging from $\Delta t=0.001$
to $\Delta t=1.0$. In this set of simulations $C_0=0.01$ and
two element orders ($10$ and $16$) have been used.
We observe that for a given spatial resolution (element order)
the computation result becomes less accurate or inaccurate
when $\Delta t$ becomes too large.
But the current scheme produces stable computations with
all these $\Delta t$ values.
In contrast, we observe that with the often-used semi-implicit
type schemes, the computation will become unstable for
moderately increased $\Delta t$ values.
In Figure \ref{fig:kovas_semiimp} we show 
the time histories of the $L^2$ norm of the $x$ component
of the velocity computed using the semi-implicit
scheme of~\cite{Dong2015clesobc} (Figure \ref{fig:kovas_semiimp}(a))
and the current scheme (Figure \ref{fig:kovas_semiimp}(b)).
These results correspond to an element order $10$, and
for the current scheme also $C_0=0.01$ in the simulations.
The computation using the semi-implicit scheme blows up
when $\Delta t$ is beyond about $0.01$, while the current scheme
exhibits a different behavior and produces stable computations
even with much larger $\Delta t$ values.


\begin{figure}
  \centerline{
    \includegraphics[width=3in]{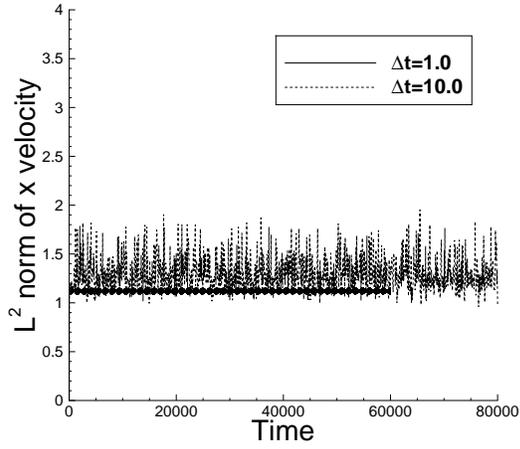}(a)
    \includegraphics[width=3in]{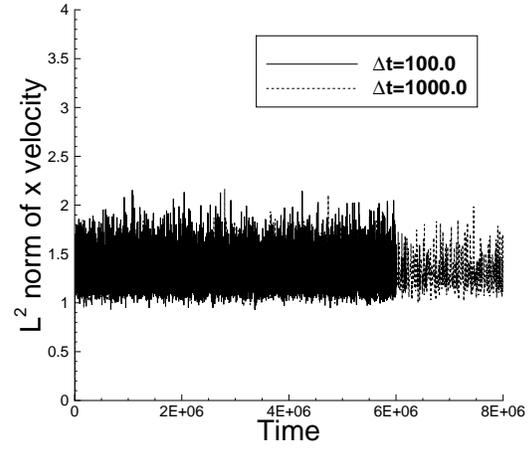}(b)
  }
  \caption{
    Kovasznay flow: Time histories of the $L^2$ norm of the $x$ velocity
    obtained using several large $\Delta t$ values:
    (a) $\Delta t=1$ and $10$,
    (b) $\Delta t=100$ and $1000$.
    Results correspond to an element order 10 and $C_0=0.01$.
  }
  \label{fig:kovas_large_dt}
\end{figure}

Figure \ref{fig:kovas_large_dt} shows the time histories of the $L^2$ norm of
the $x$ velocity of the Kovasznay flow obtained using the current scheme
with several large time step sizes ranging from
$\Delta t=1.0$ to $\Delta t=1000$. The element order is $10$ and
$C_0=0.01$ in these simulations.
While we cannot expect accuracy in these
results because of the large $\Delta t$ values, 
the computations using the current scheme
are nonetheless stable with these large $\Delta t$ 
for the Kovasznay flow.


\begin{figure}
  \centerline{
    \includegraphics[width=2.8in]{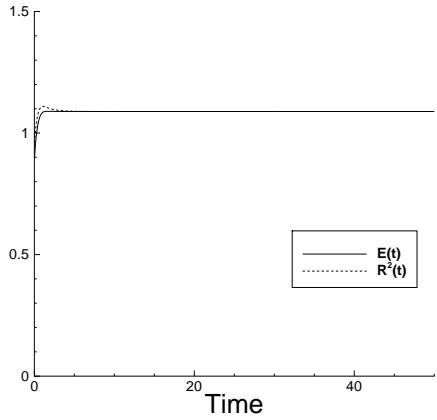}(a)
    \includegraphics[width=2.8in]{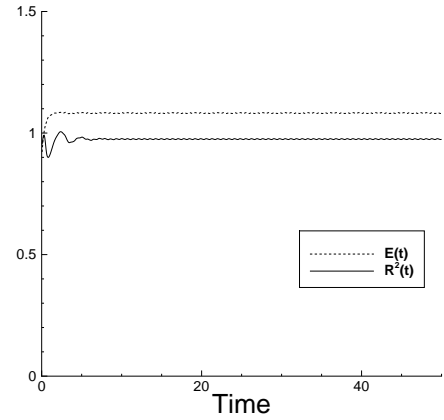}(b)
  }
  \centerline{
    \includegraphics[width=2.8in]{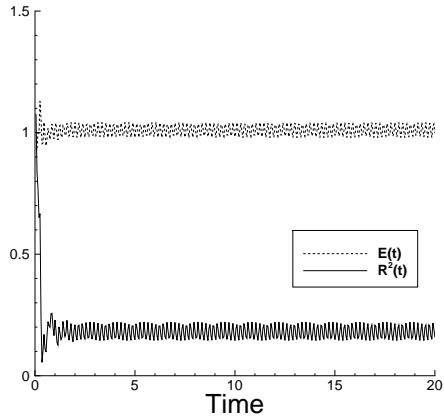}(c)
    \includegraphics[width=2.8in]{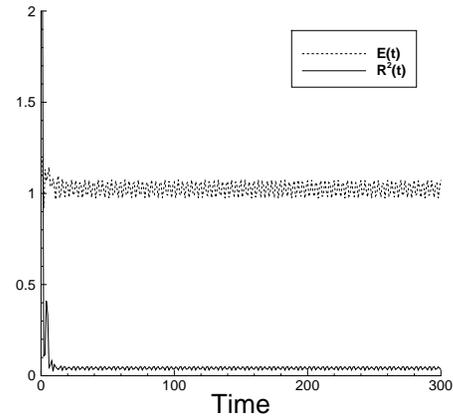}(d)
  }
  \caption{
    Kovasznay flow: time histories of $E(t)$ and $R^2(t)$ obtained
    using several time step sizes:
    (a) $\Delta t=0.005$, (b) $\Delta t=0.01$,
    (c) $\Delta t=0.05$, (d) $\Delta t=1.0$.
    Results correspond to $C_0=0.01$ and an element order $10$
    in the simulations.
  }
  \label{fig:R_E_hist_kovas}
\end{figure}

In Figure \ref{fig:R_E_hist_kovas} we plot the time histories of
two variables: $E(t)$ defined by \eqref{equ:def_energy},
and $[R(t)]^2$ computed by solving \eqref{equ:R_equ_mod}
from the current scheme. These two variables are
supposed to be equal due to equation \eqref{equ:def_R}.
With small $\Delta t$ values, the two quantities resulting
from the computations are indeed the same, which is
evident from the time histories obtained
with $\Delta t=0.005$ in Figure \eqref{fig:R_E_hist_kovas}(a).
The initial difference between the curves for $E(t)$ and $R^2(t)$ in
Figure \eqref{fig:R_E_hist_kovas}(a) is due to the initial
condition, because the zero initial velocity field is not compatible
with the Dirichlet boundary condition (non-zero velocity)
on the domain boundary.
When $\Delta t$ becomes moderately large or very large, we
can observe a difference between the computed $E(t)$ and $R^2(t)$,
and the discrepancy between them becomes larger
with increasing $\Delta t$ (Figures \ref{fig:R_E_hist_kovas}(b)-(d)).
Both $E(t)$ and $R^2(t)$ fluctuate over time
in the simulations using large $\Delta t$ values.
While the computed $E(t)$ approximately stays at a constant mean level,
$R(t)$ appears to be driven toward zero with very large $\Delta t$
in the simulations (Figure \ref{fig:R_E_hist_kovas}(d)).
%
These results suggest that when $\Delta t$ becomes large
the dynamical system consisting of equations \eqref{equ:alg_1}--\eqref{equ:alg_6}
seems to be able to automatically adjust
the $R(t)$ and the $\frac{R(t)}{\sqrt{E(t)}}$ levels.
The term $\frac{R(t)}{\sqrt{E(t)}}$ places a control
on the explicitly-treated nonlinear term in the Navier-Stokes
equation \eqref{equ:alg_1}. With very large $\Delta t$ values,
the system drives $\frac{R(t)}{\sqrt{E(t)}}$ toward zero, thus
making the computation more stable or stabilizing the computation.
This seems to be the mechanism by which
the current scheme produces stable simulations
with large $\Delta t$ for the Kovasznay flow.


\subsection{Flow Past a Circular Cylinder in a Periodic Channel}
\label{sec:cyl}

In this section
we investigate the flow past a circular cylinder inside a periodic
channel, which is driven by a constant pressure gradient
along the channel direction. We employ this problem to test
the algorithm developed herein.


\begin{figure}
  \centering
  \includegraphics[height=1.6in]{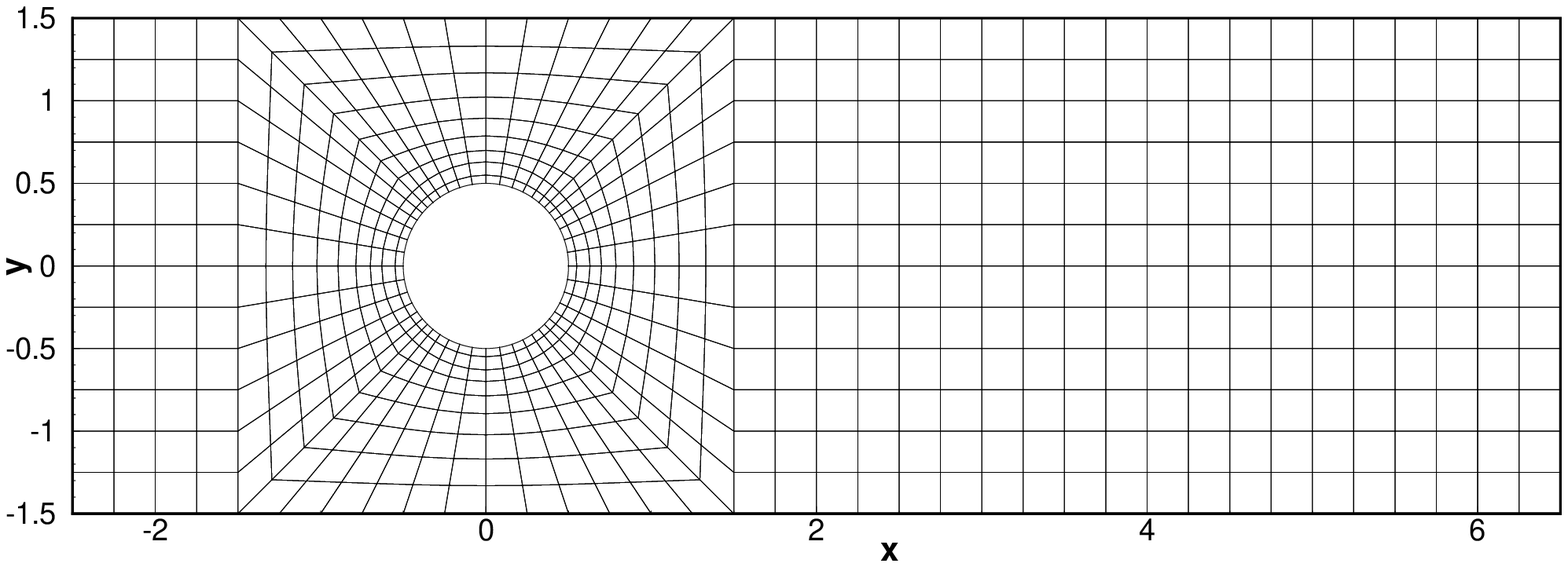}(a)
  \includegraphics[height=1.6in]{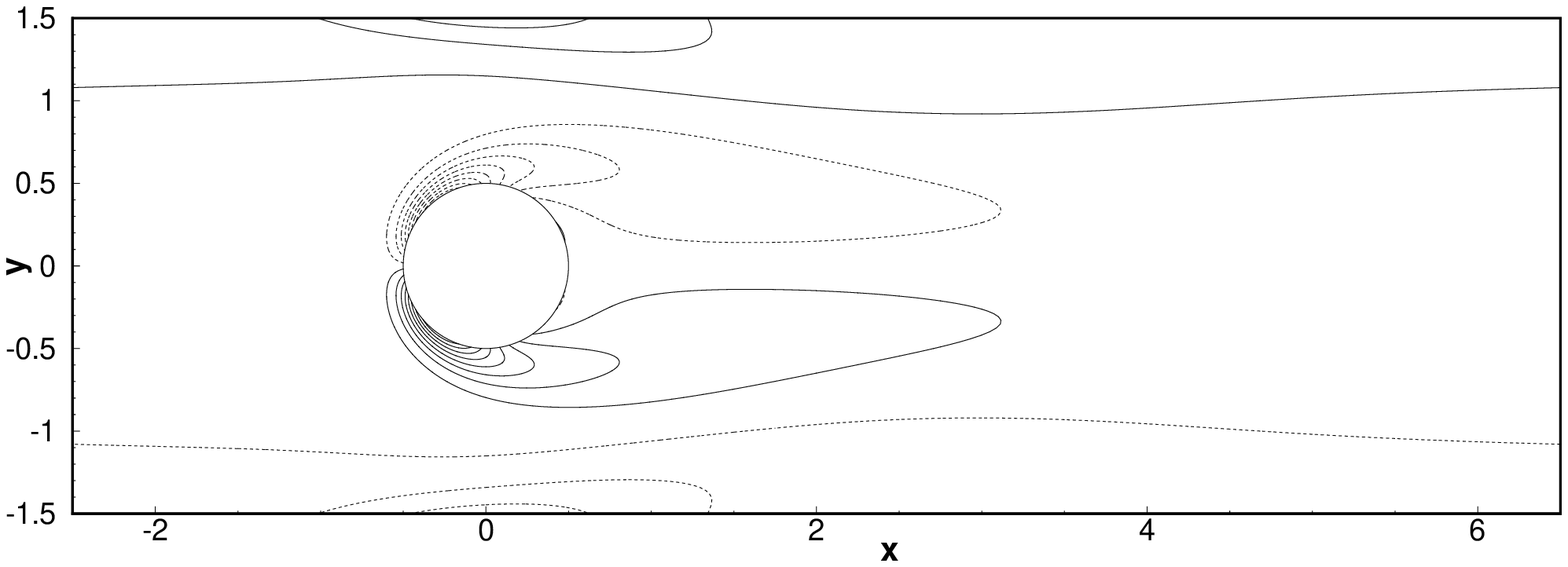}(b)
  \includegraphics[height=1.6in]{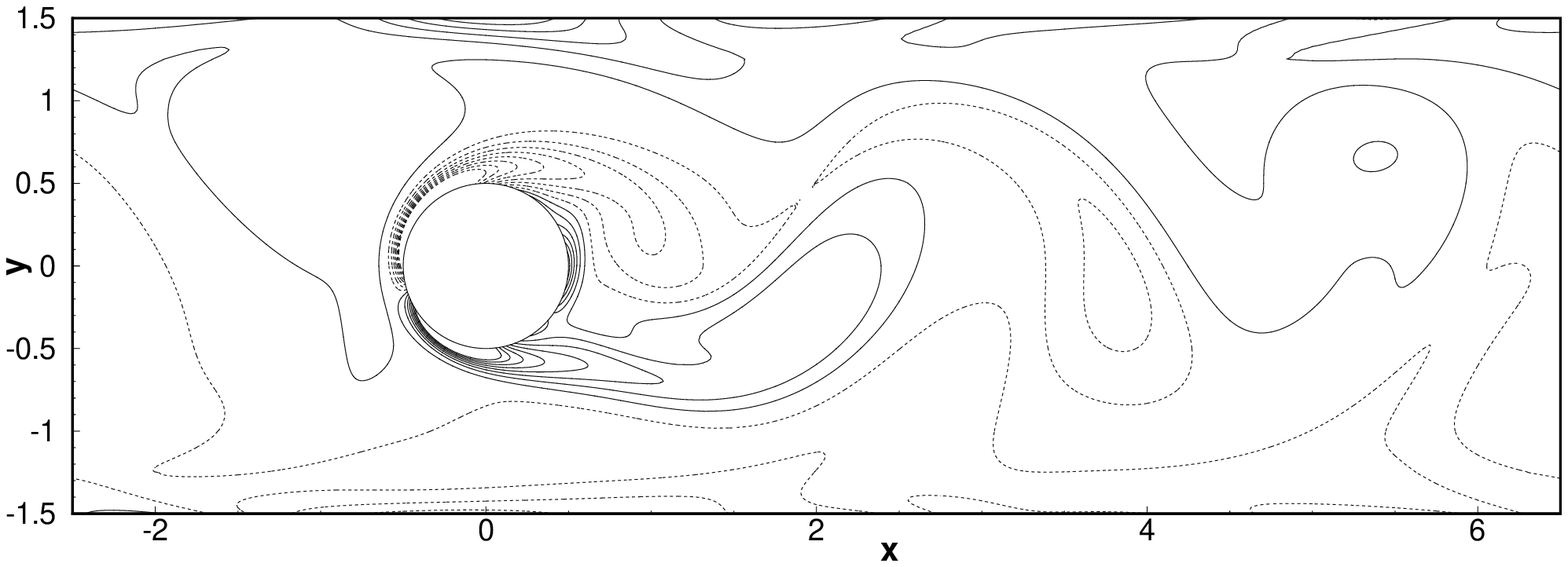}(c)
  \caption{
    Cylinder flow in a periodic channel:
    (a) Mesh of $720$ quadrilateral elements.
    Vorticity contours at $\nu=0.01$ (b) and $\nu=0.005$ (c), with the flow
    driven by a normalized pressure gradient of $0.02$.
  }
  \label{fig:cyl_mesh}
\end{figure}

Consider a circular cylinder of diameter $D$ placed inside a horizontal channel
occupying the domain $-2.5D\leqslant x\leqslant 6.5D$ and
$-1.5D\leqslant y\leqslant 1.5D$, as depicted in Figure \ref{fig:cyl_mesh}(a).
The top and bottom of the domain ($y=\pm 1.5D$) are
the channel walls. A pressure gradient is imposed along the $x$ direction
to drive the flow.
The flow domain and all the physical variables
are assumed to be periodic in the horizontal direction
at $x=-2.5D$ and $x=6.5D$.
The center of the cylinder coincides with the origin
of the coordinate system.
This setting is equivalent to the flow past an infinite
sequence of circular
cylinders inside
an infinitely long horizontal channel.


We use the cylinder diameter $D$ as the characteristic length
scale. Let $\frac{\Delta P}{D}$ denote the pressure gradient (body force)
that drives the flow along the $x$ direction, and let
$g_0$ denote a unit body force magnitude. Then the non-dimensional body
force in equation \eqref{equ:nse} has the magnitude
$|\mathbf{f}| = \frac{\Delta P}{g_0D}$ and points along
the $x$ direction. We use $U_0=\sqrt{\frac{g_0D}{\rho_f}}$,
where $\rho_f$ is the fluid density, as the characteristic
velocity scale. All the length variables are normalized by
$D$, and all velocity variables are normalized by $U_0$.

Figure \ref{fig:cyl_mesh}(a) shows the spectral element mesh
used to discretize the domain, which consists of $720$
quadrilateral elements. The element order has been varied between
$2$ and $6$ in the simulations to test the effect of spatial
resolutions on the simulation results.
On the top and bottom channel walls, as well as on the cylinder
surface, a no-slip condition is imposed on the velocity,
that is, the boundary condition \eqref{equ:bc_vel}
with $\mathbf{w}=0$. In the horizontal direction (at $x/D=-2.5$ and $6.5$)
periodic conditions are imposed for all the flow variables.
Long-time simulations have been performed using the
algorithm from Section \ref{sec:method} for a range of 
Reynolds numbers (or $\nu$) and the driving force $|\mathbf{f}|$.
The simulation is started at a low Reynolds number with
a zero initial velocity field at the very beginning. Then
the Reynolds number is increased incrementally. A snapshot of
the flow field from a lower Reynolds number is used as
the initial condition in the simulation of the next larger Reynolds number.
At each Reynolds number,
a long-time simulation is performed (with typically 
$100$ flow-through times), such that the flow has reached a statistically
stationary state for that Reynolds number and the initial condition
will have no effect on the flow. 
A range of values for $\Delta t$ and $C_0$ have been tested in
the simulations.


An overview of the flow features is provided by
Figures \ref{fig:cyl_mesh}(b) and (c), in which the contours of
the instantaneous vorticity have been shown for
two Reynolds numbers corresponding to $\nu=0.01$ and $\nu=0.005$
and a non-dimensional driving pressure gradient $|\mathbf{f}|=0.02$.
The dashed curves denote negative vorticity values.
With $\nu=0.01$ the flow reaches a steady state eventually.
The pattern of the vorticity contours around the cylinder is
typical of the cylinder flow~\cite{DongKER2006} at low Reynolds numbers.
Due to presence of the channel, we can also observe a certain level of
vorticity distribution
near the upper/lower channel walls above or below the cylinder. 
With $\nu=0.005$ an unsteady flow is observed, with
 periodical vortex shedding from the cylinder into the wake. 
 Because of the periodicity in the horizontal direction,
 the vortices shed to the cylinder wake will re-enter the domain
 on the left side and influence the flow
 upstream of the cylinder. These upstream vortices
 interact with the cylinder, leading to more complicated dynamics
 and flow structures in the cylinder wake.

%

\begin{table}
  \begin{center}
    \begin{tabular}{lllllll}
      \hline
      $\nu$ & element order & mean-drag & rms-drag & mean-lift & rms-lift & driving force \\ \hline
      $0.01$ & $2$ & $0.487$ & $0$ & $0$ & $0$ & $0.524$ \\
      & $3$ & $0.527$ & $0$ & $0$ & $0$ & $0.524$ \\
      & $4$ & $0.524$ & $0$ & $0$ & $0$ & $0.524$ \\
      & $5$ & $0.524$ & $0$ & $0$ & $0$ & $0.524$ \\
      & $6$ & $0.524$ & $0$ & $0$ & $0$ & $0.524$ \\
      \hline
      $0.001$ & $2$ & $0.353$ & $0.158$ & $-8.90e-4$ & $0.270$ & $0.524$ \\
      & $3$ & $0.519$ & $0.269$ & $-1.35e-3$ & $2.83e-2$ & $0.524$ \\
      & $4$ & $0.530$ & $0.277$ & $-3.72e-4$ & $1.36e-2$ & $0.524$ \\
      & $5$ & $0.526$ & $0.280$ & $1.62e-5$ & $3.80e-3$ & $0.524$ \\
      & $6$ & $0.524$ & $0.289$ & $8.96e-5$ & $1.64e-3$ & $0.524$ \\
      \hline
    \end{tabular}
  \end{center}
  \caption{
    Effect of spatial resolution on the computed forces acting on channel/cylinder
    walls. Drag refers to the force in the $x$ direction and lift refers
    to the force in the $y$ direction. Driving force is the normalized total force driving
    the flow due to the imposed pressure gradient.
  }
  \label{tab:force_resolution}
\end{table}

The effect of the spatial mesh resolution on the simulation results
is demonstrated by Table \ref{tab:force_resolution}.
We have computed the total forces acting on the walls
(sum of those on the cylinder surface and channel wall surfaces)
from the simulations.
This table lists the time-averaged mean drag (or drag if steady flow),
root-mean-square (rms) drag, mean lift, and rms lift on the walls
 corresponding
to a non-dimensionalized driving pressure gradient
$\frac{\Delta P}{g_0D}=0.02$.
Several element orders,
ranging from $2$ to $6$, are tested in the simulations.
The results for two Reynolds numbers corresponding to
$\nu=0.01$ and $\nu=0.001$ have been included in the table, and they
are obtained using $\Delta t=0.01$ in the simulations.
The total driving force on the flow is 
$
\frac{\Delta P}{g_0D} \times V_{\Omega}
= 0.02\times 26.2146
\approx 0.524,
$
where $V_{\Omega}=((6.5+2.5)\times(1.5+1.5) - \pi/4)\approx 26.2146$
is the normalized volume (or area) of the flow domain.
At steady state or statistically stationary state of the flow,
this driving force will be balanced by the total force exerted on the flow by
the walls. This will give rise to a total mean drag
acting on the walls with the same value as
the total driving force.
Therefore, we expect that the time-averaged
mean drag (or the drag if steady flow) obtained
from the simulations should be equal or close to the total driving force
in the flow domain. This can be used to check the accuracy of the simulations.
With very low element orders, we observe a discrepancy between the computed
drag from the simulations and the expected value.
As the element order increases in the simulations, this discrepancy
decreases and becomes negligible or completely vanishes.
For $\nu=0.01$, the computed drag becomes very close to the expected value
for element order $3$, and with element orders $4$ and above the computed
drag matches the expected value.
For $\nu=0.001$, the difference between the computed mean-drag and
the expected value becomes very small with element orders $5$ and above.


\begin{figure}
  \centerline{
    \includegraphics[width=1.8in]{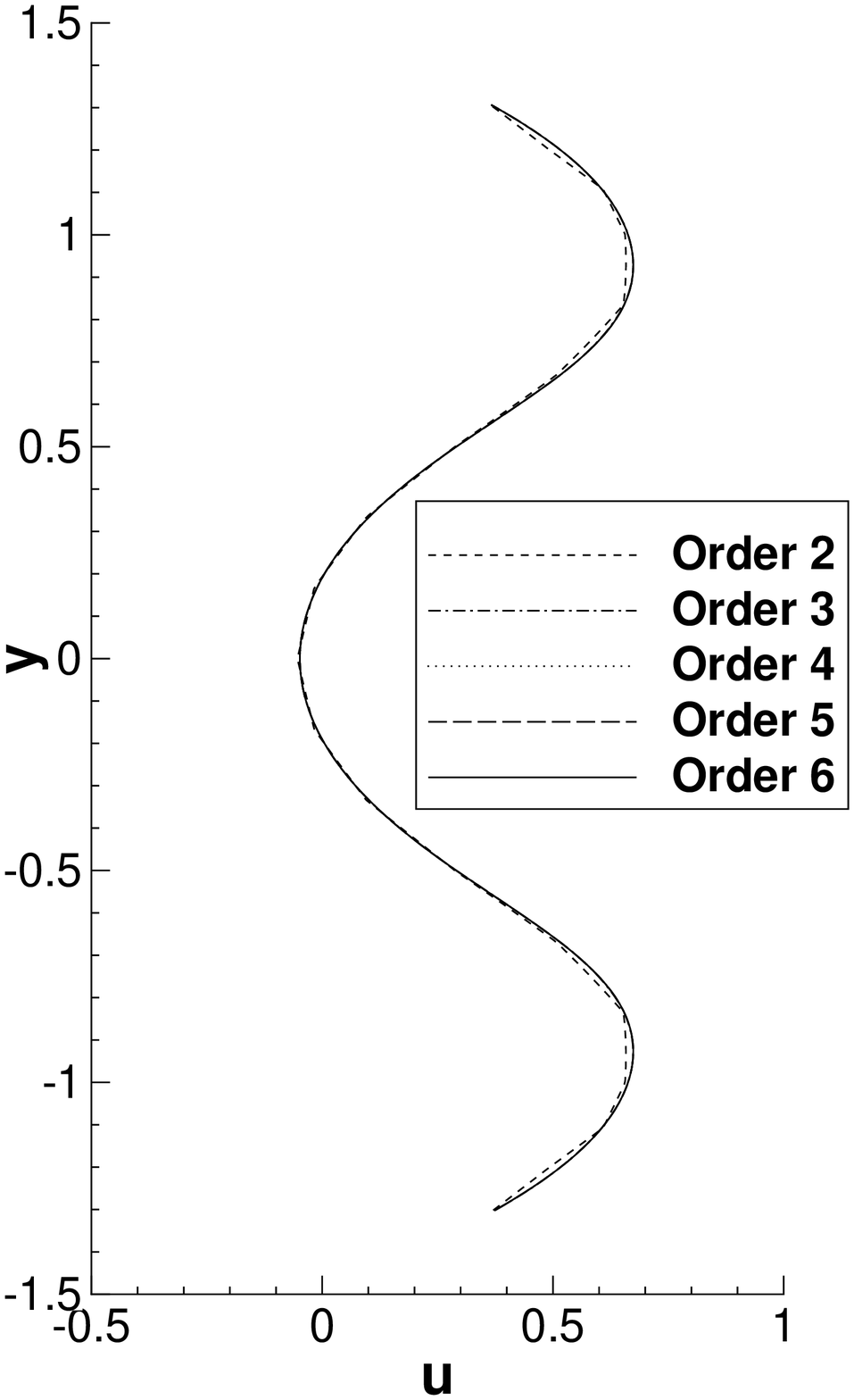}(a)
    \includegraphics[width=1.8in]{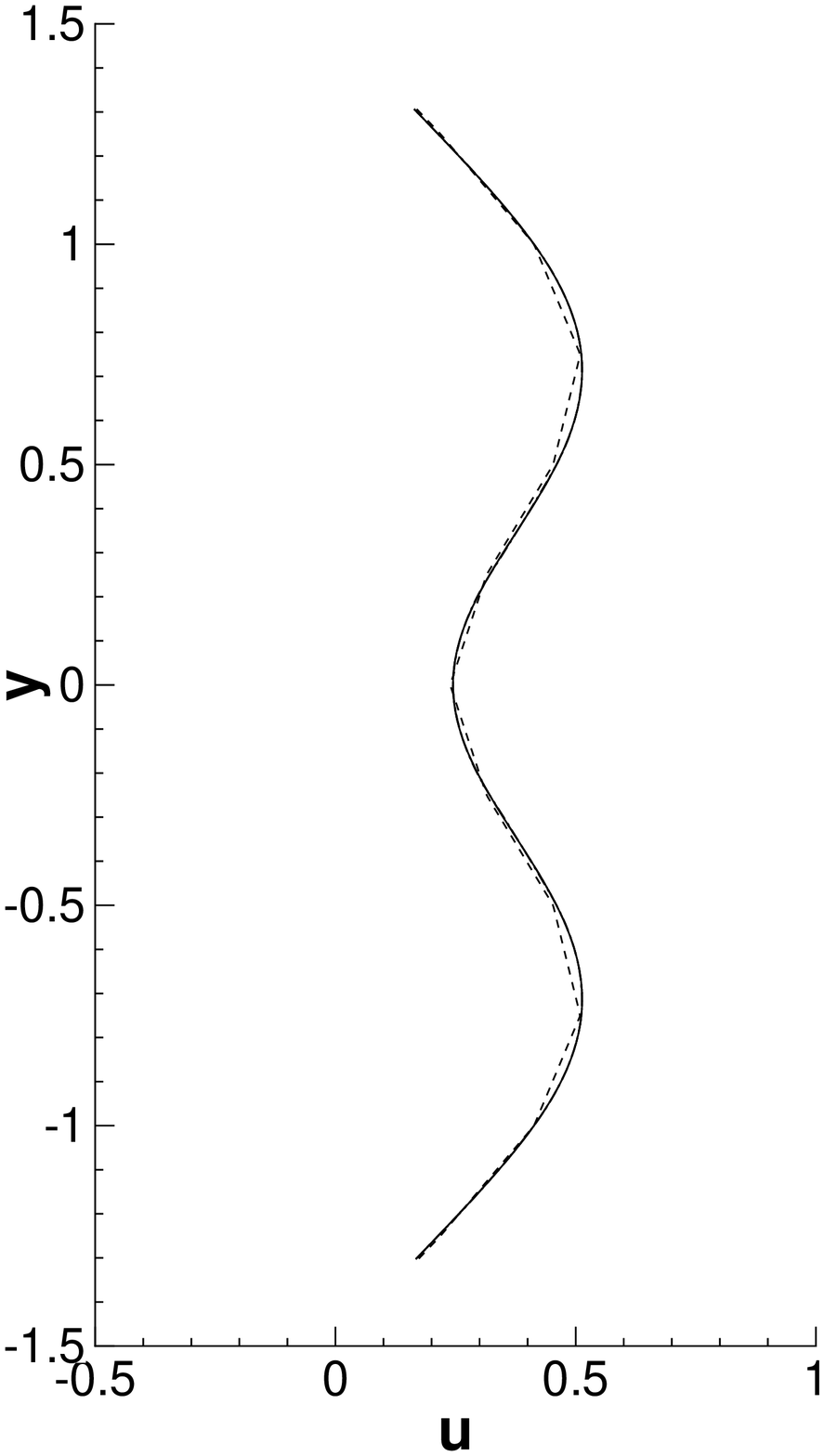}(b)
    \includegraphics[width=1.8in]{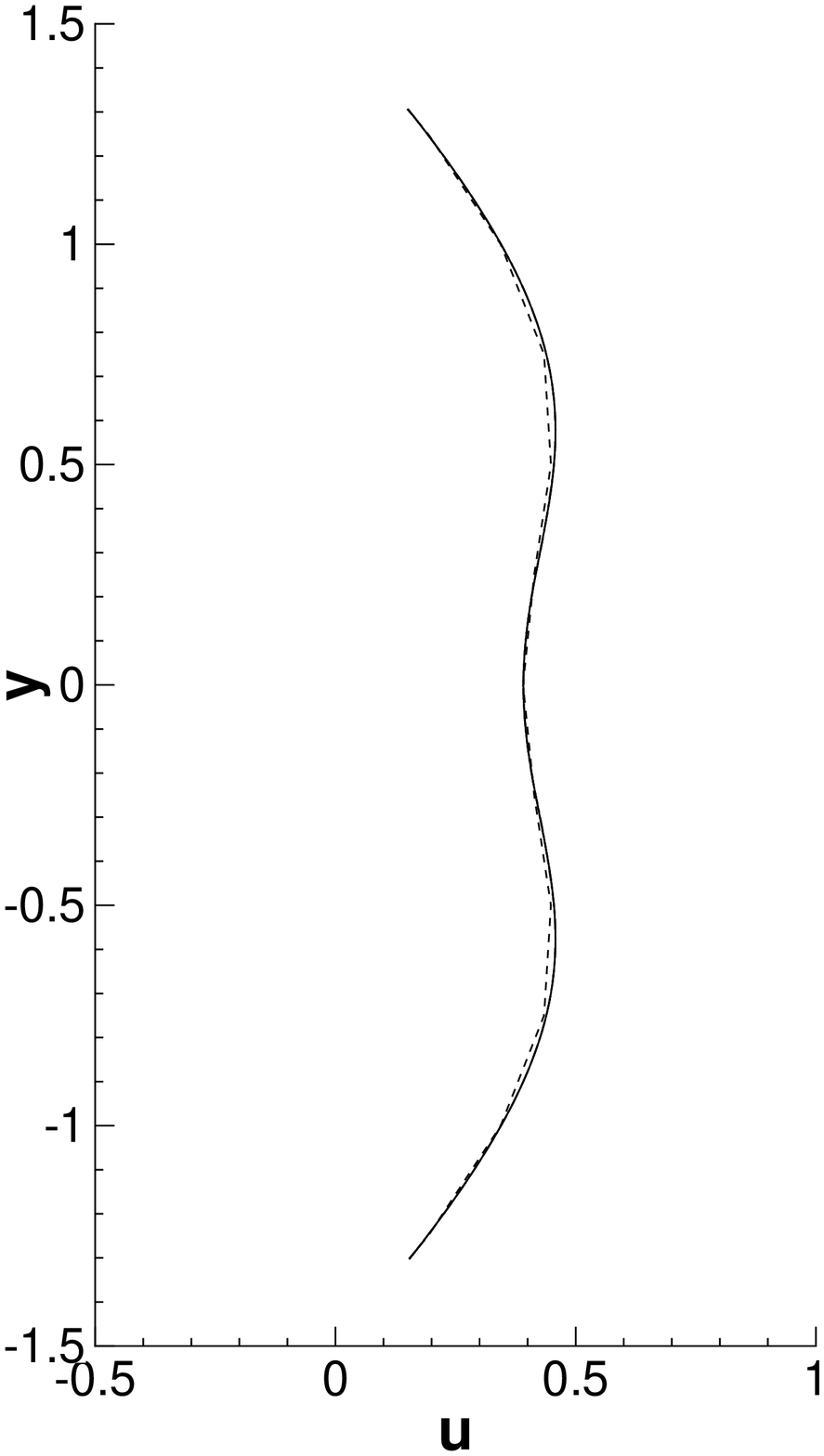}(c)
  }
  \centerline{
    \includegraphics[width=4.5in]{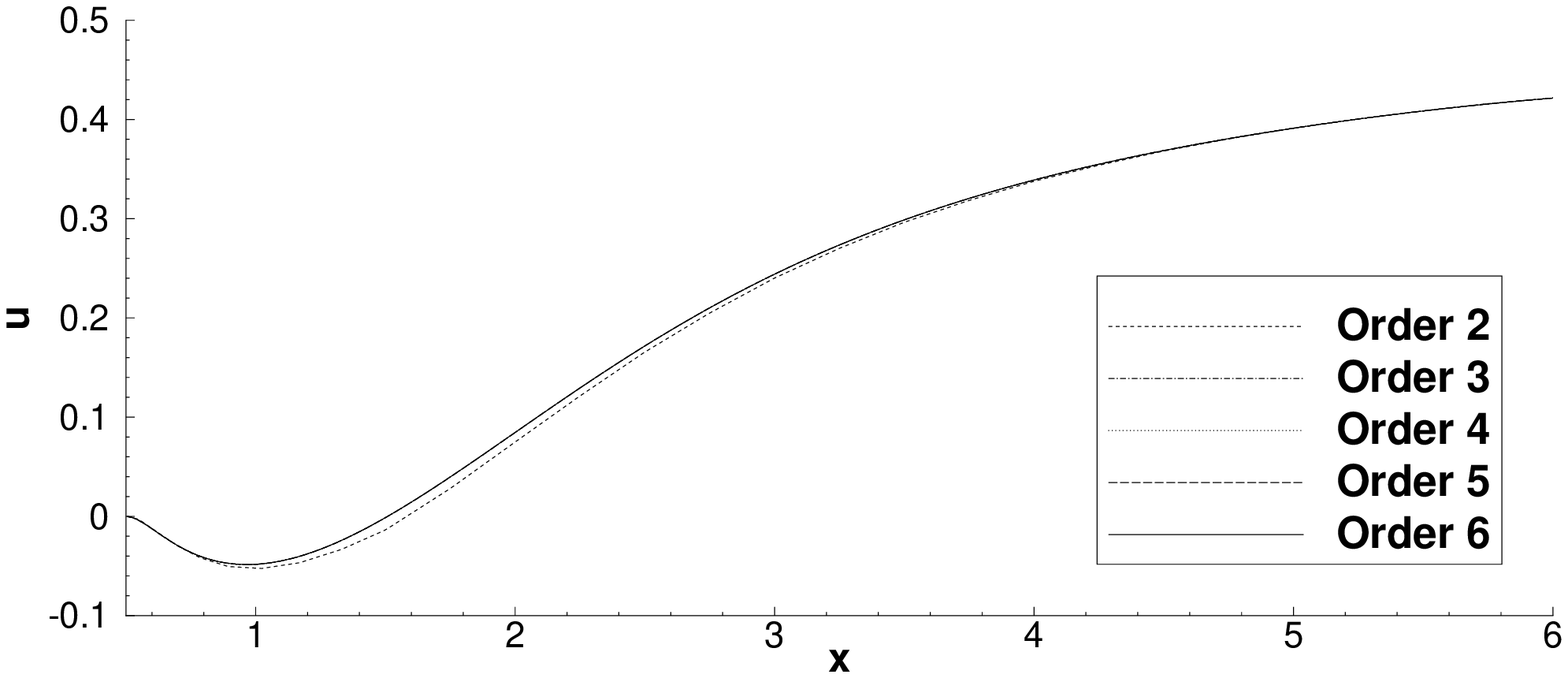}(d)
  }
  \caption{
    Comparison of stream-wise velocity profiles ($\nu=0.01$) computed using 
    different element orders at downstream locations
    (a) $x/d=1.0$, (b) $x/d=3.0$, (c) $x/d=5.0$
    and the velocity profiles along the centerline of domain
    (d) $y=0.0$.
    Results are obtained with $\Delta t=0.01$ and $C_0=1000$ in
    the simulations.
  }
  \label{fig:comp_vel_profile}
\end{figure}

Figure \ref{fig:comp_vel_profile} shows a comparison of the
profiles of the steady-state stream-wise velocity along the cross-flow direction
at several downstream locations $x/D=1$, $3$, and $5$,
as well as along the centerline of the domain ($y=0$),
for the Reynolds number corresponding to $\nu=0.01$ and with
several element orders in the simulations.
The total driving force in the domain is $0.524$.
While some difference can be observed between the profile
corresponding to element order $2$ and the other profiles,
all the velocity profiles obtained with the element orders $3$
and above basically overlap with one another, suggesting the
independence with respect to the spatial resolutions.
In light of these observations, the majority of simulations
reported below are performed using an element order $4$, and
for higher Reynolds numbers the results corresponding to an element order
$5$ are also employed in the simulations.


\begin{figure}
  \centerline{
    \includegraphics[height=3in]{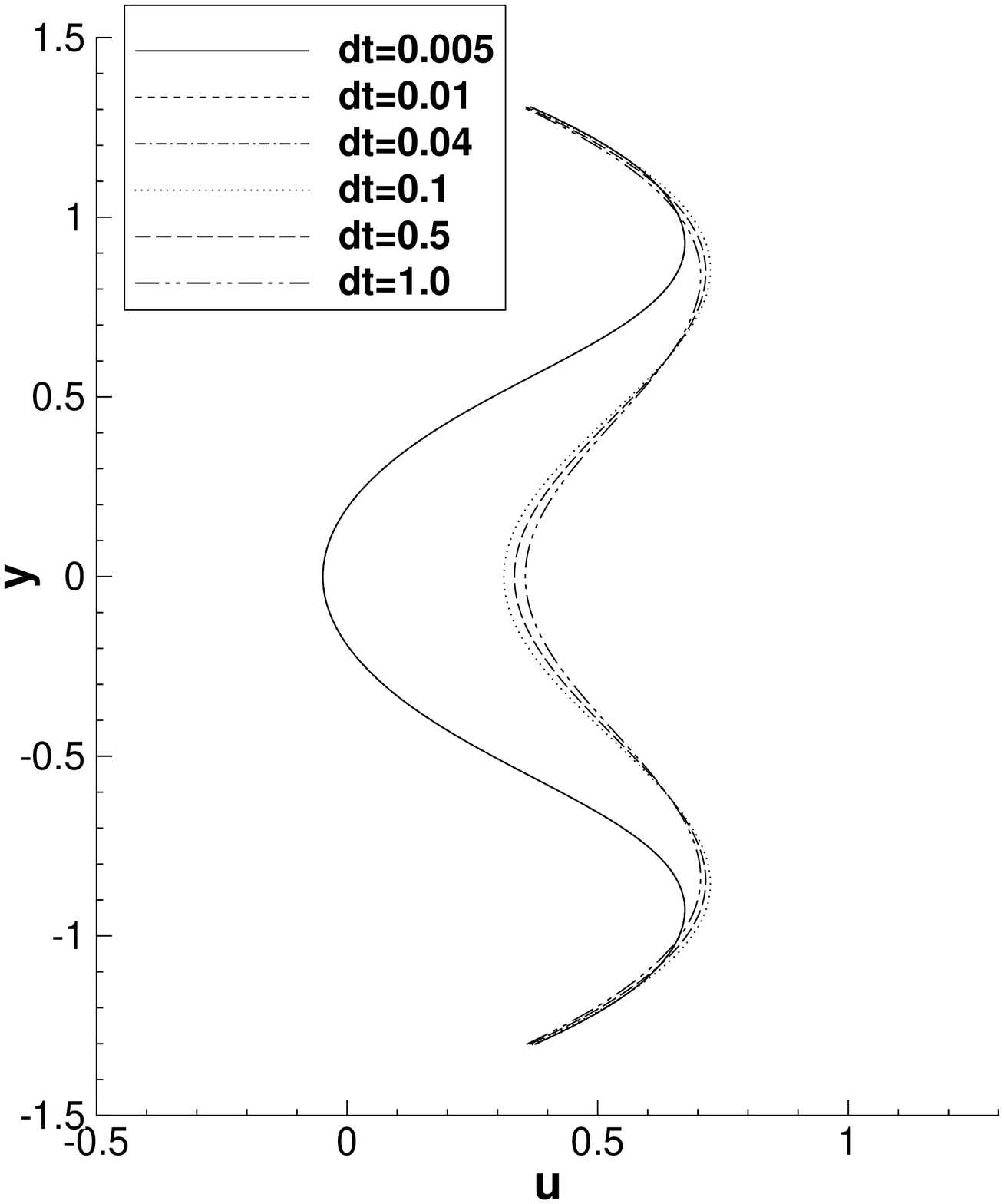}(a)
    \includegraphics[height=3in]{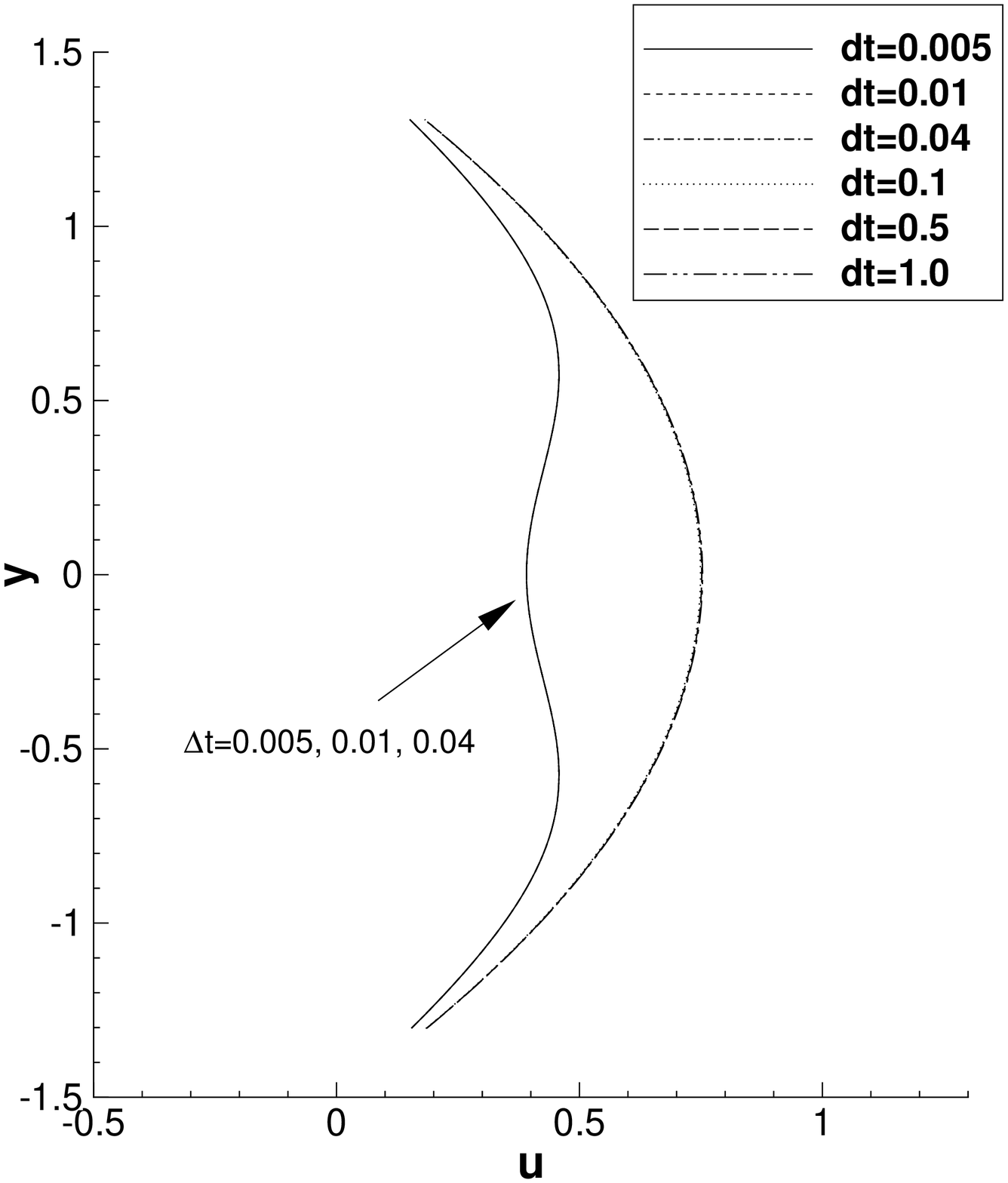}(b)
  }
  \caption{
    Effect of the time step size $\Delta t$ ($\nu=0.01$): stream-wise velocity profiles
    at downstream locations (a) $x/D=1.0$ and (b) $x/D=5.0$ obtained
    with different time step sizes.
  }
  \label{fig:dt_effect}
\end{figure}

The effect of the time step size $\Delta t$ on the simulation
results is illustrated by Figure \ref{fig:dt_effect},
in which we compare the stream-wise velocity profiles
along the $y$ direction at the downstream locations $x/D=1$ and $5$
obtained with time step sizes ranging from
$\Delta t=0.005$ to $\Delta t=1.0$.
These results are computed with an element order $4$
for the Reynolds number corresponding to $\nu=0.01$,
and the total driving force is $0.524$ in the domain.
The profiles obtained with time step sizes $\Delta t=0.04$ and smaller
all overlap with one another. 
With larger $\Delta t$ values (e.g.~$\Delta t=0.1$, $0.5$ and $1.0$),
our method is also able to produce stable simulations at this
Reynolds number. But the obtained velocity profiles exhibit a pronounced
difference when compared with those computed using smaller $\Delta t$
values, indicating that these results are no longer accurate. 
From numerical experiments we observe that
for steady-state flow problems the current method seems to be able to produce
stable computations, even with very large $\Delta t$ values in the simulations,
as have been shown here and in  Section \ref{sec:kovas}.
For Reynolds numbers at which the flow is unsteady,
despite the energy stability property (Theorem \ref{thm:thm_1}),
we observe that in practice there is a restriction on the maximum $\Delta t$
that can be used in the simulations, at least with our current implementation
of the scheme. The computation will become unstable when $\Delta t$ exceeds this value.
For the results reported subsequently, the simulations are performed
with $\Delta t=0.01$ for lower Reynolds numbers ($\nu\geqslant 0.005$),
and $\Delta t=0.001$ 
for higher Reynolds numbers ($\Delta t=5E-4$ for
the Reynolds number corresponding to $\nu=2\times 10^{-4}$).


\begin{figure}
  \centerline{
    \includegraphics[height=2.5in]{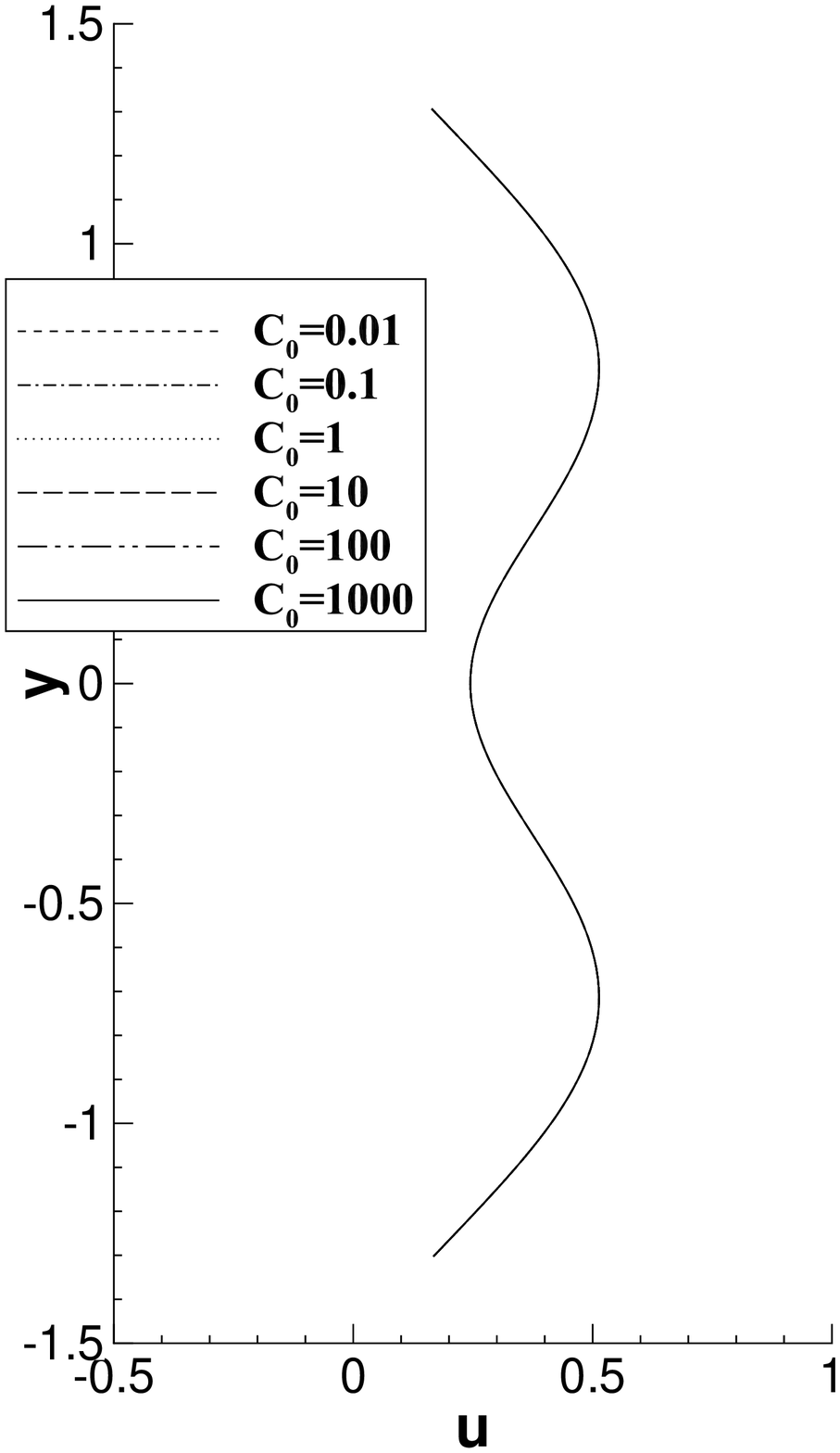}(a)
    \includegraphics[height=2.3in]{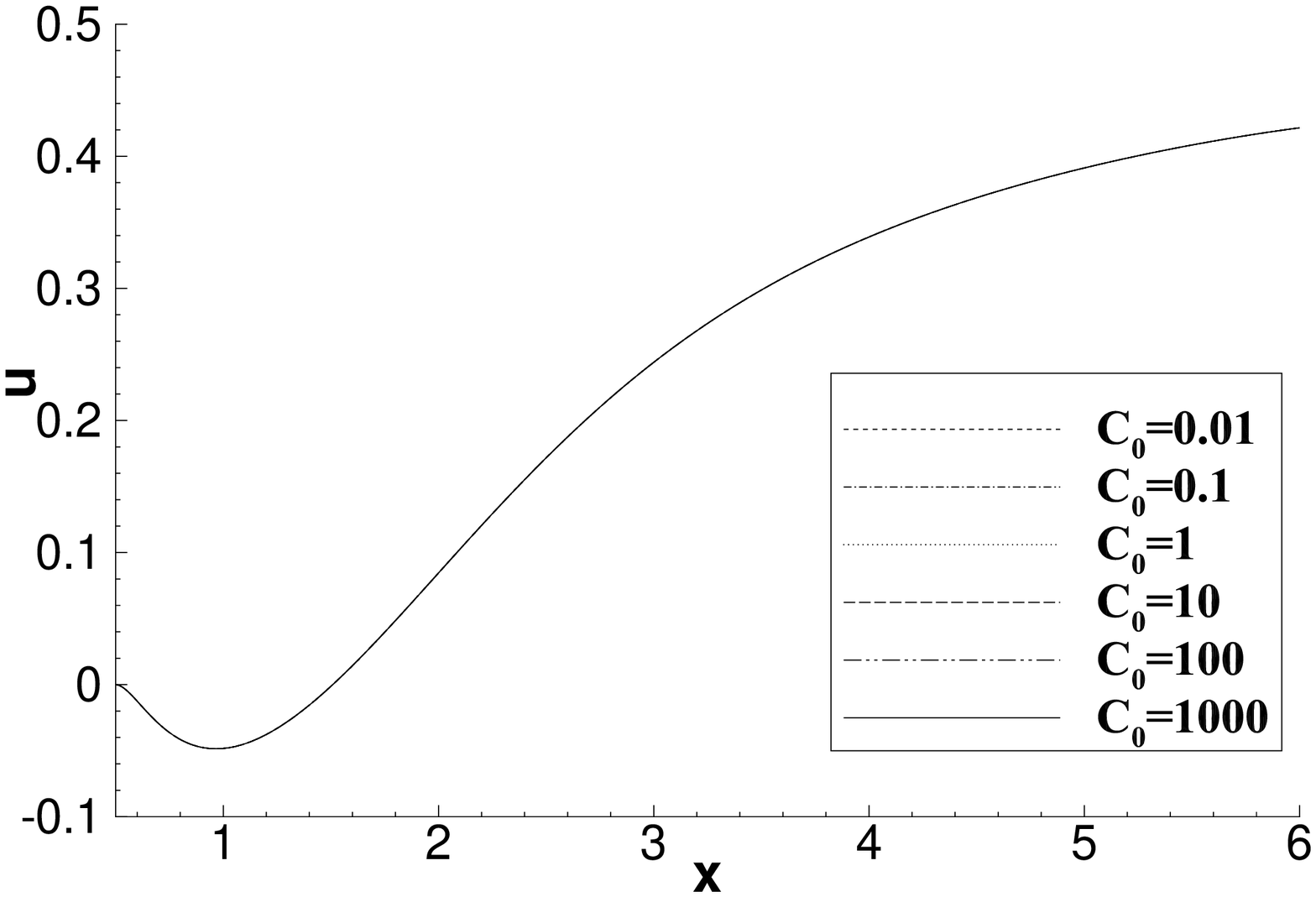}(b)
  }
  \caption{
    $C_0$ effect:
    Comparison of stream-wise velocity profiles ($\nu=0.01$)
    at (a) $x=3.0$ and (b) along the centerline, obtained
    using a range of $C_0$ values in the algorithm.
  }
  \label{fig:vel_profile_C0}
\end{figure}

The constant $C_0$ in the current algorithm has been observed to
influence the accuracy of the steady-state solution for the Kovasznay flow
in the previous section.
Whether $C_0$ affects the accuracy of results for the current problem
has also been studied, and we observe no apparent effect of the $C_0$ value on
the forces and the flow field distributions.
For the Reynolds number corresponding
to $\nu=0.01$ and a total driving force $0.524$, we have performed simulations
with the $C_0$ constant ranging from $C_0=0.01$ to $C_0=1000$
in the algorithm. The computed forces on the cylinder/channel walls
from all these simulations match the total force imposed to drive the flow. 
Figure \ref{fig:vel_profile_C0} compares the stream-wise velocity profiles
at the downstream location $x/D=3$ and along the centerline of the
domain ($y=0$) computed using different $C_0$ values in
the current algorithm. The results correspond to an
element order $4$ and time step size $\Delta t=0.01$ in the simulations.
The velocity profiles for different cases exactly overlap with
one another, suggesting that the computed velocity field is not sensitive
to $C_0$ in the simulations.
In the discussions of subsequent results, we employ a value
$C_0=1000$ in the simulations unless otherwise specified.


\begin{table}
  \begin{center}
    \begin{tabular}{l l l l l l l}
      \hline
      $\nu$ & mean-drag & rms-drag & mean-lift & rms-lift & Pressure Gradient & Driving Force \\ \hline
      $0.01$ & $0.131$ & $0$ & $0$ & $0$ & $0.005$ & $0.131$ \\
      & $0.262$ & $0$ & $0$ & $0$ & $0.01$ & $0.262$ \\
      & $0.524$ & $0$ & $0$ & $0$ & $0.02$ & $0.524$ \\
      & $0.787$ & $0$ & $0$ & $0$ & $0.03$ & $0.786$ \\
      & $1.049$ & $0$ & $0$ & $0$ & $0.04$ & $1.049$ \\
      & $1.311$ & $0$ & $0$ & $0$ & $0.05$ & $1.311$ \\
      \hline
      $0.002$ & $0.131$ & $0.0105$ & $9.64e-5$ & $3.17e-4$ & $0.005$ & $0.131$ \\
      & $0.264$ & $0.0519$ & $1.97e-5$ & $8.79e-4$ & $0.01$ & $0.262$ \\
      & $0.529$ & $0.150$ & $-9.10e-5$ & $3.77e-3$ & $0.02$ & $0.524$ \\
      & $0.793$ & $0.277$ & $8.85e-5$ & $8.82e-3$ & $0.03$ & $0.786$ \\
      & $1.059$ & $0.427$ & $-4.77e-4$ & $1.45e-2$ & $0.04$ & $1.049$ \\
      & $1.324$ & $0.574$ & $2.35e-4$ & $2.22e-2$ & $0.05$ & $1.311$ \\
      \hline
    \end{tabular}
  \end{center}
  \caption{
    Comparison of the
    forces on the cylinder/channel walls from the simulations and the imposed
    pressure gradient and total driving force. 
  }
  \label{tab:force_varied}
\end{table}

We have varied the magnitude of the driving pressure gradient
systematically ranging from $\frac{\Delta P}{g_0D}=0.005$
to $0.05$, and carried out simulations corresponding
to each of the force values. In Table \ref{tab:force_varied} we 
list the drag and lift on the walls obtained from the simulations
for two fixed Reynolds numbers corresponding to $\nu=0.01$
and $\nu=0.002$. The imposed pressure gradient and
the total driving force in the flow domain have also been
shown. These results are obtained with an element order $4$ and
a time step size $\Delta t=0.01$ in the simulations.
Note that physically the mean drag from the simulations is expected to match the 
imposed total driving force.
At the lower Reynolds number ($\nu=0.01$) the computed mean drag
on the walls is essentially the same as the total driving force
for different cases.
At the higher Reynolds number ($\nu=0.002$) the computed mean-drag values
also agree well with the driving forces. The discrepancy is less than
$1\%$ for the range of driving forces considered here.
The mean lift is zero or essentially zero, and the
rms lift is also observed to be small.



\begin{figure}
  \centering
  \includegraphics[width=4.5in]{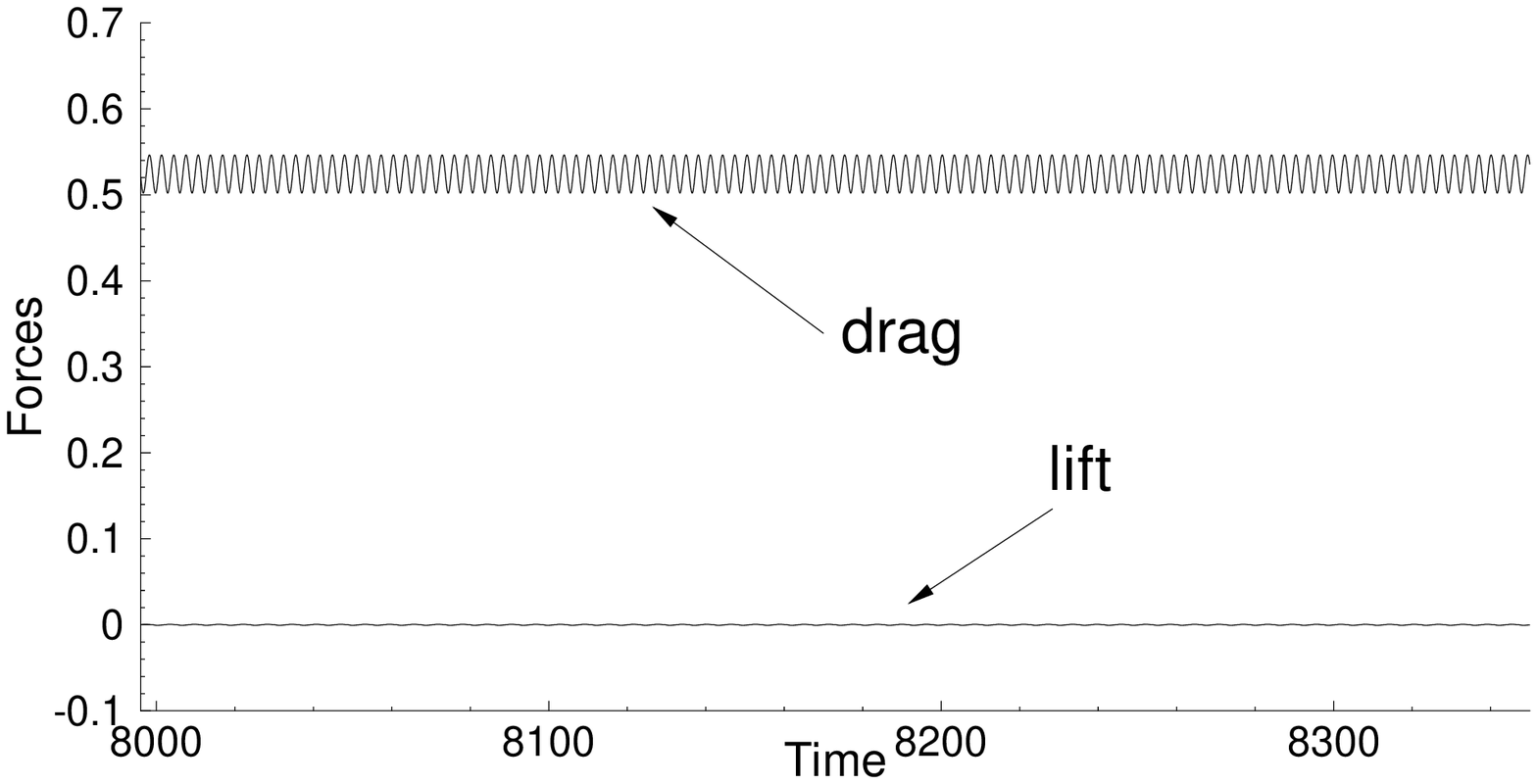}(a)
  \includegraphics[width=4.5in]{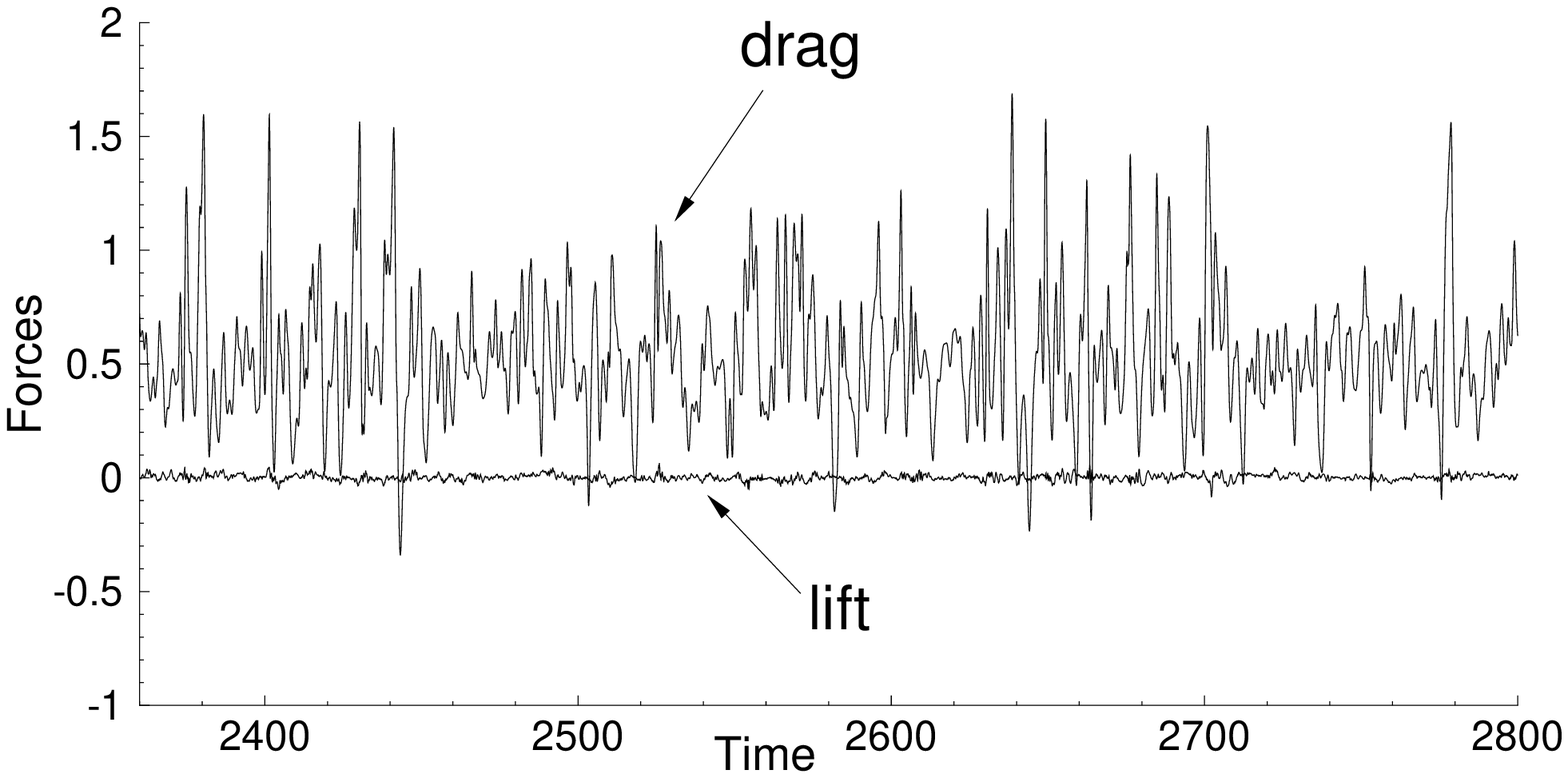}(b)
  \caption{
    Time histories of the drag and lift on the cylinder/channel walls for 
    (a) $\nu=0.005$ and (b) $\nu=0.001$.
    The total driving force in the domain is $0.524$.
  }
  \label{fig:force_hist_nu}
\end{figure}

Figure \ref{fig:force_hist_nu} shows a window of the time histories of
the drag and lift on the domain walls at two Reynolds numbers
corresponding to $\nu=0.005$ and $\nu=0.001$, with a non-dimensional driving force
$0.524$ in the domain.
Periodic vortex shedding behind the cylinder can be observed at
$\nu=0.005$ (see Figure \ref{fig:cyl_mesh}(c) for the vorticity distribution),
resulting in a periodic drag signal fluctuating about a constant mean value.
As the viscosity decreases, vortices shed to the cylinder wake
persist downstream and
can be observed to re-enter the domain through the left boundary.
The interactions between these upstream vortices and the cylinder
cause the vortex shedding from the cylinder and
the forces acting on the walls to become
highly irregular. At $\nu=0.001$ one can observe
chaotic fluctuations in the time histories of the drag and lift
on the channel/cylinder walls (Figure \ref{fig:force_hist_nu}(b)).
Because of the confinement effect of the channel, the total
lift acting on the domain walls appears quite insignificant, and it does not
exhibit the large fluctuations as typically observed in
the flow past a cylinder in an open domain
(see e.g.~\cite{DongKER2006,DongTK2008}).


\begin{table}
  \begin{center}
    \begin{tabular}{l l l l l l}
      \hline
      $\nu$ & mean-drag & rms-drag & mean-lift & rms-lift & driving force \\ \hline
      $0.02$ & $0.524$ & $0.0$ & $0.0$ & $0.0$ & $0.524$ \\
      $0.01$ & $0.524$ & $0.0$ & $0.0$ & $0.0$ & $0.524$ \\
      $0.005$ & $0.525$ & $0.0158$ & $-5.30e-8$ & $3.87e-4$ & $0.524$ \\
      $0.00333$ & $0.526$ & $0.0760$ & $2.69e-4$ & $1.00e-3$ & $0.524$ \\
      $0.002$ & $0.529$ & $0.150$ & $-9.10e-5$ & $3.77e-3$ & $0.524$ \\
      $0.001$ & $0.530$ & $0.277$ & $-3.72e-4$ & $1.36e-2$ & $0.524$ \\
      $0.0005$ & $0.524$ & $0.403$ & $3.67e-4$ & $3.73e-2$ & $0.524$ \\
      $0.0002$ & $0.523$ & $0.590$ & $-6.85e-4$ & $4.35e-2$ & $0.524$ \\
      \hline
    \end{tabular}
  \end{center}
  \caption{
    Forces on cylinder/channel walls
    obtained with different fluid viscosities.
    Simulations are performed with an element order $5$ for $\nu=0.0002$ and
    $4$ for the other $\nu$ values.
  }
  \label{tab:force_viscosity}
\end{table}

We have simulated this flow problem for a range of Reynolds numbers corresponding to
viscosities ranging from $\nu=0.02$ to $\nu=0.0002$, with
the total driving force fixed at $0.524$ (corresponding to
a non-dimensional pressure gradient $0.02$).
Table \ref{tab:force_viscosity} lists the mean and rms forces acting on
the walls obtained from the simulations corresponding to different
Reynolds numbers. In the simulations the element
order is $5$ for the case $\nu=0.0002$
and $4$ for the other cases. 
As expected, the mean drag values obtained
from the simulations are very close to or the same as the driving force for
different Reynolds numbers, indicating that the method has captured
the flow quite accurately.
The rms drag is observed to increase significantly with increasing
Reynolds number (decreasing $\nu$), while the rms lift remains
insignificant for the range of Reynolds numbers considered here.


\begin{figure}
  \centerline{
    \includegraphics[width=3.1in]{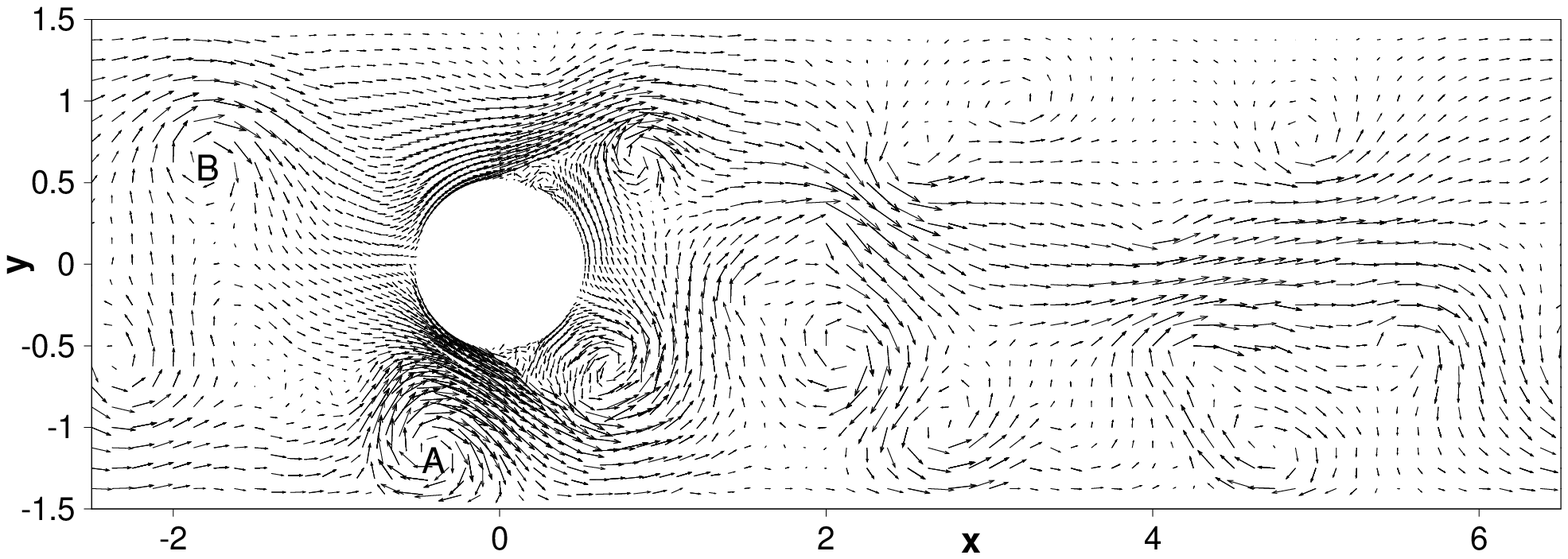}(a)
    \includegraphics[width=3.1in]{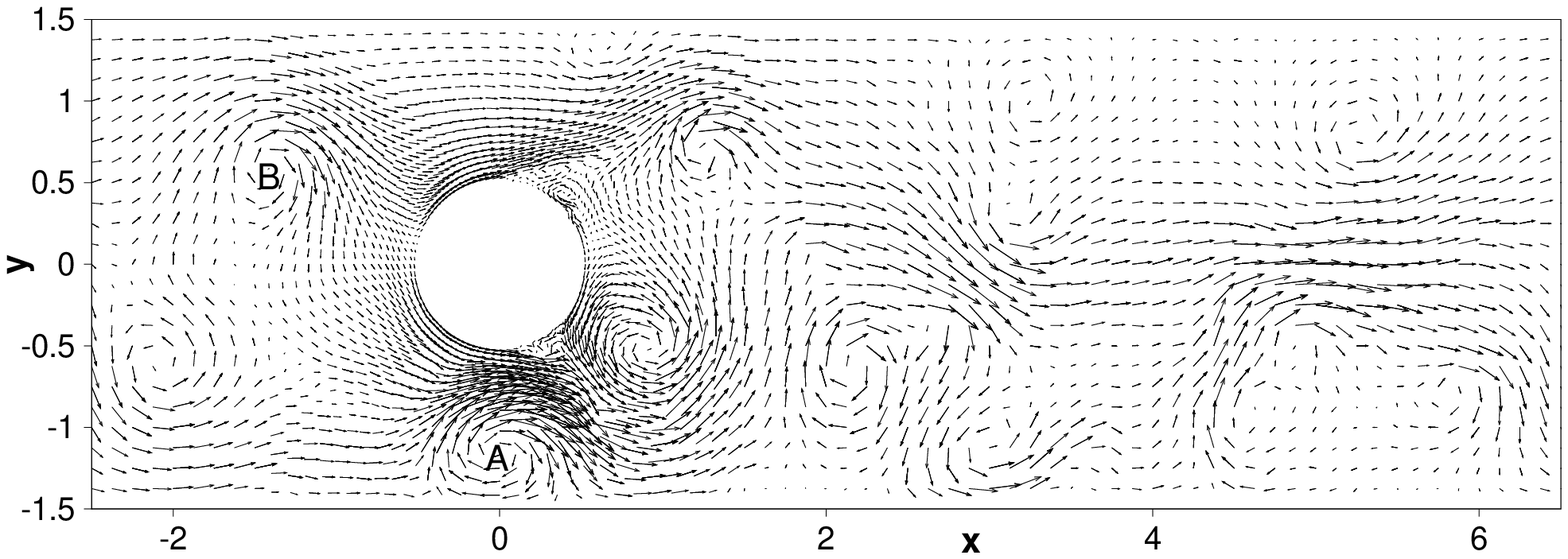}(b)
  }
  \centerline{
    \includegraphics[width=3.1in]{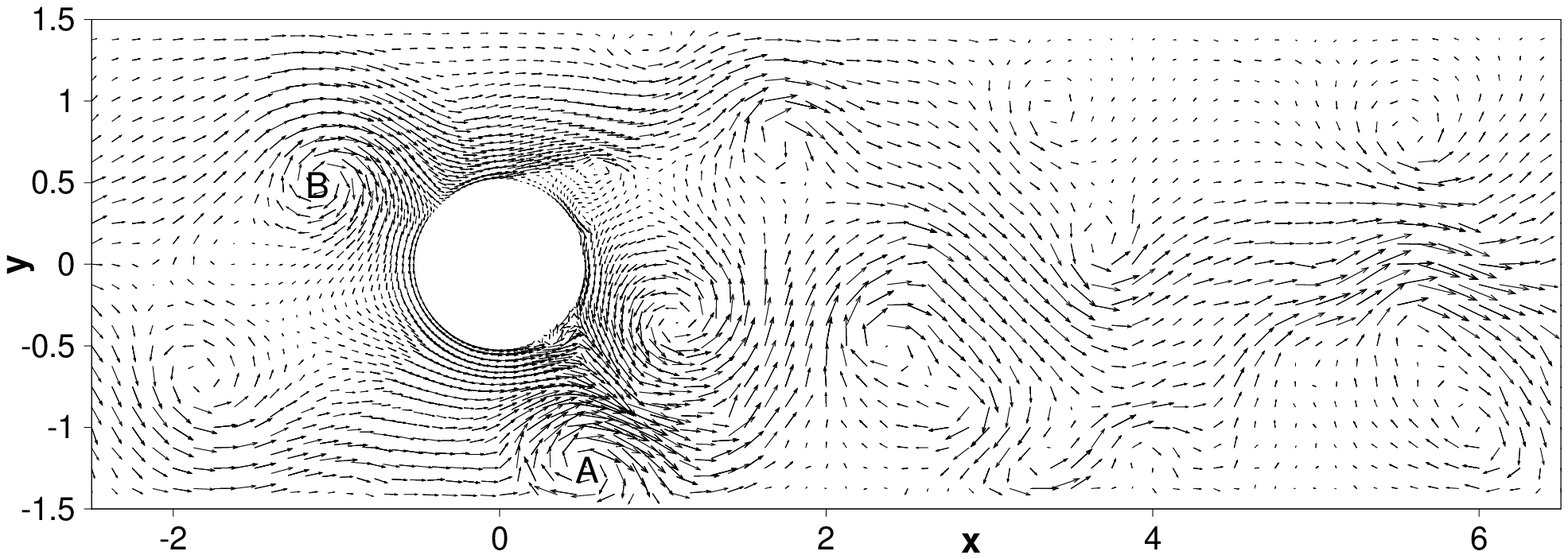}(c)
    \includegraphics[width=3.1in]{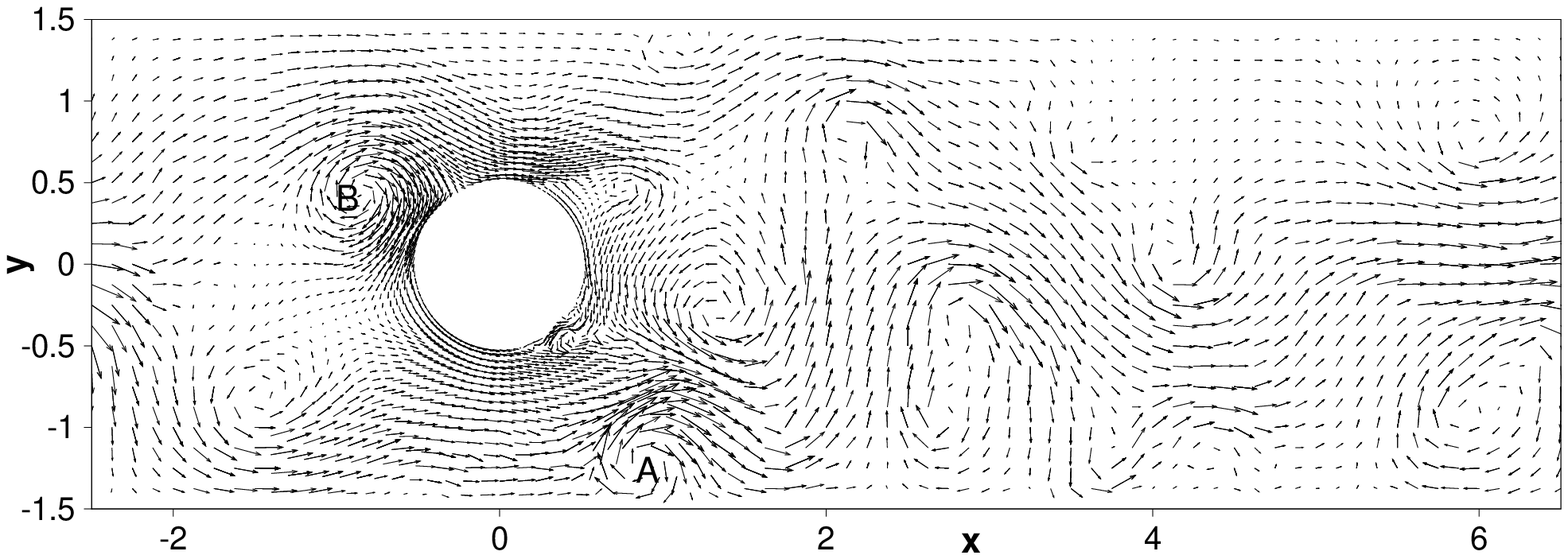}(d)
  }
  \centerline{
    \includegraphics[width=3.1in]{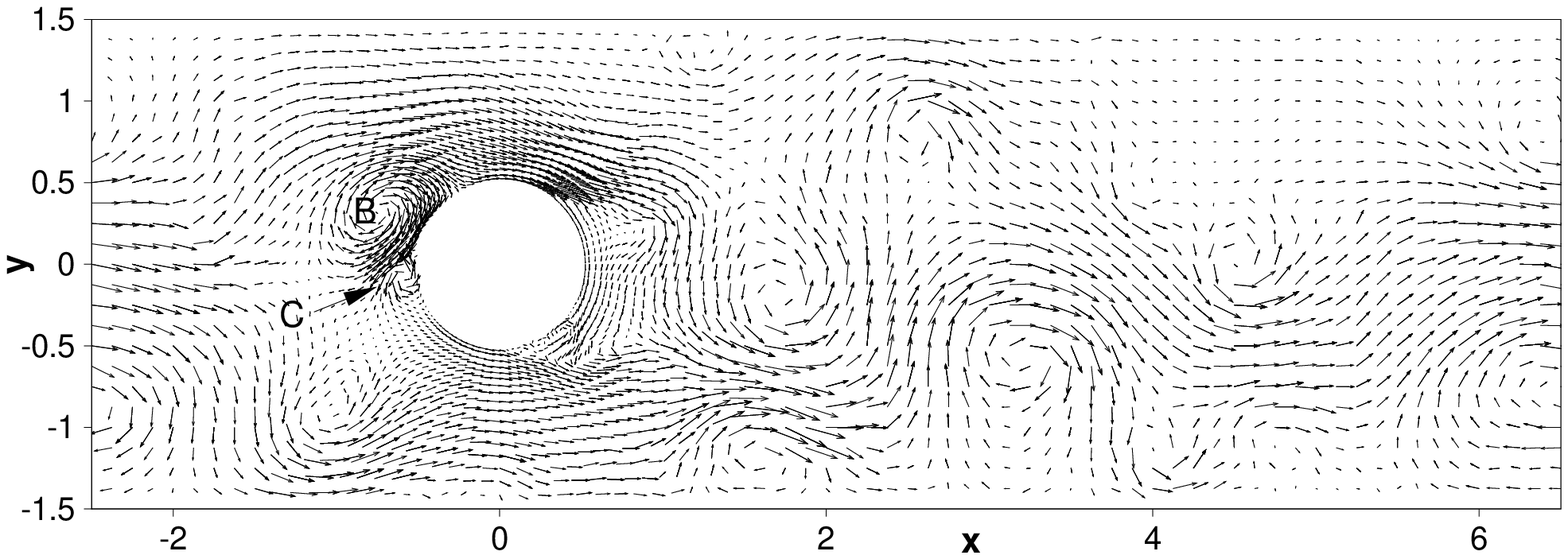}(e)
    \includegraphics[width=3.1in]{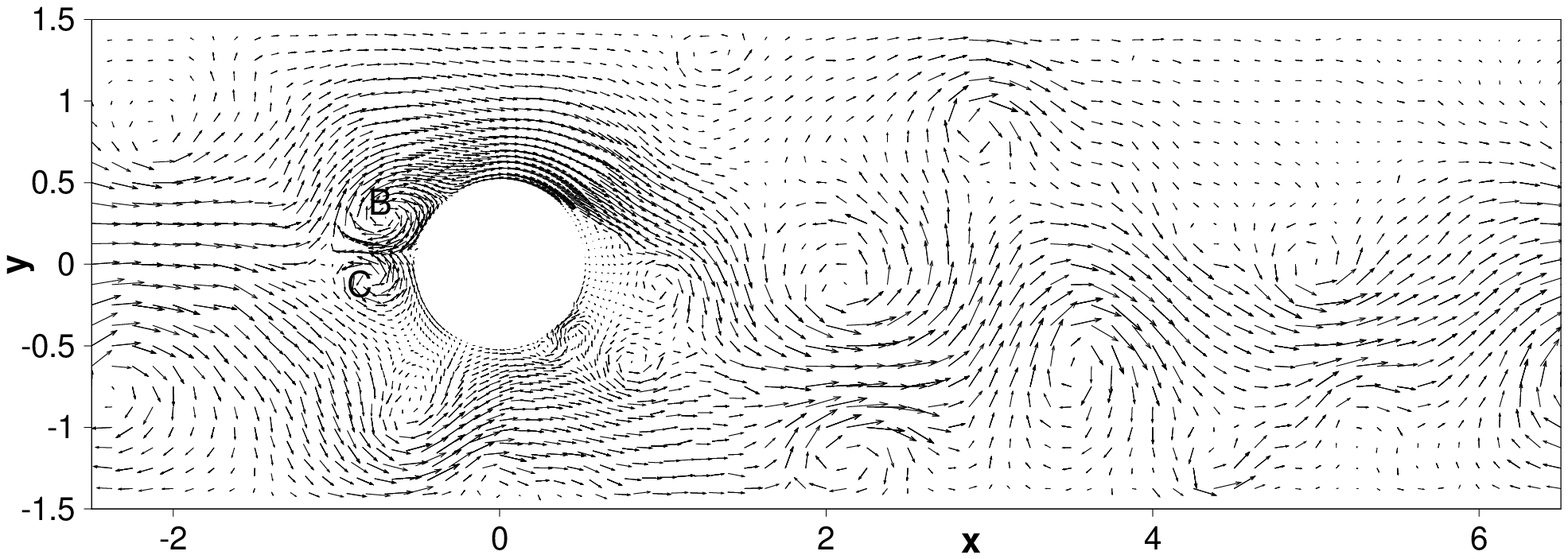}(f)
  }
  \centerline{
    \includegraphics[width=3.1in]{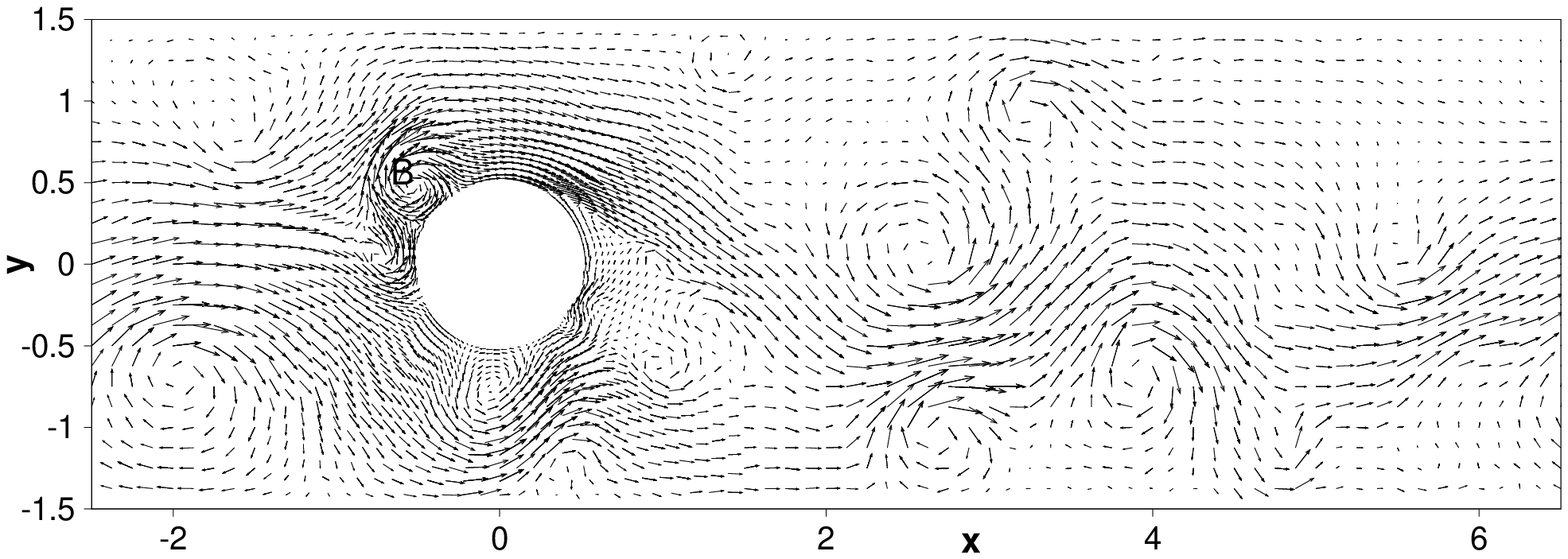}(g)
    \includegraphics[width=3.1in]{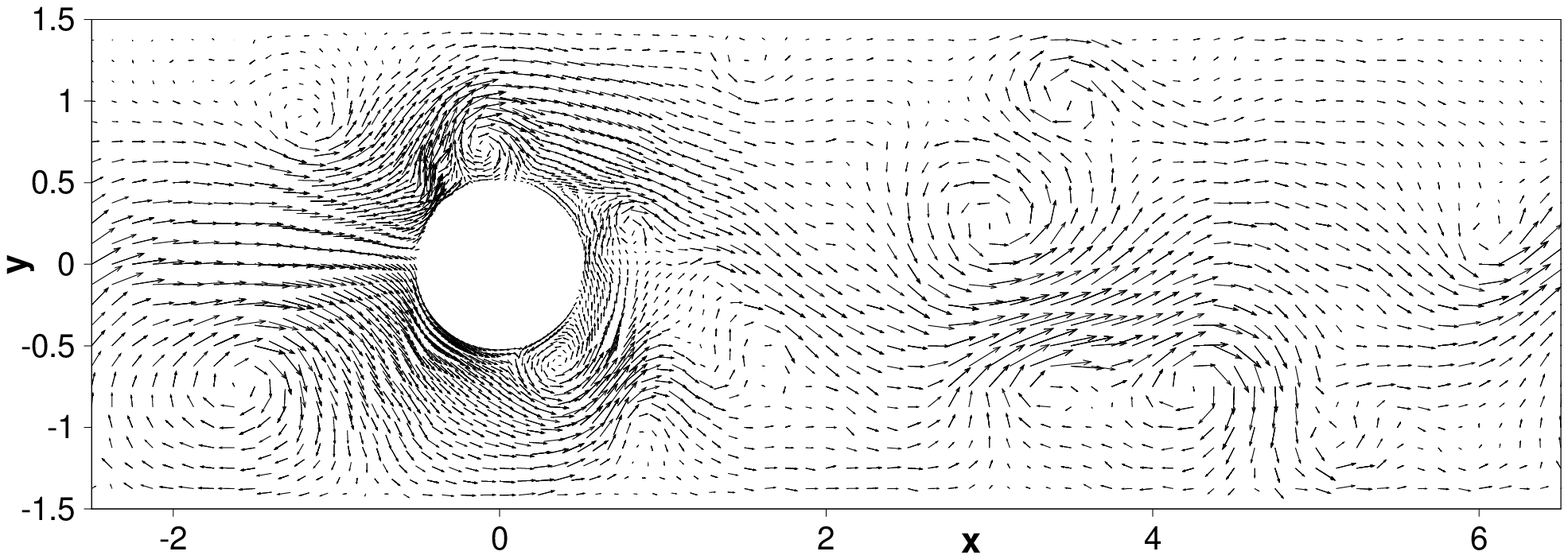}(h)
  }
  \centerline{
    \includegraphics[width=3.1in]{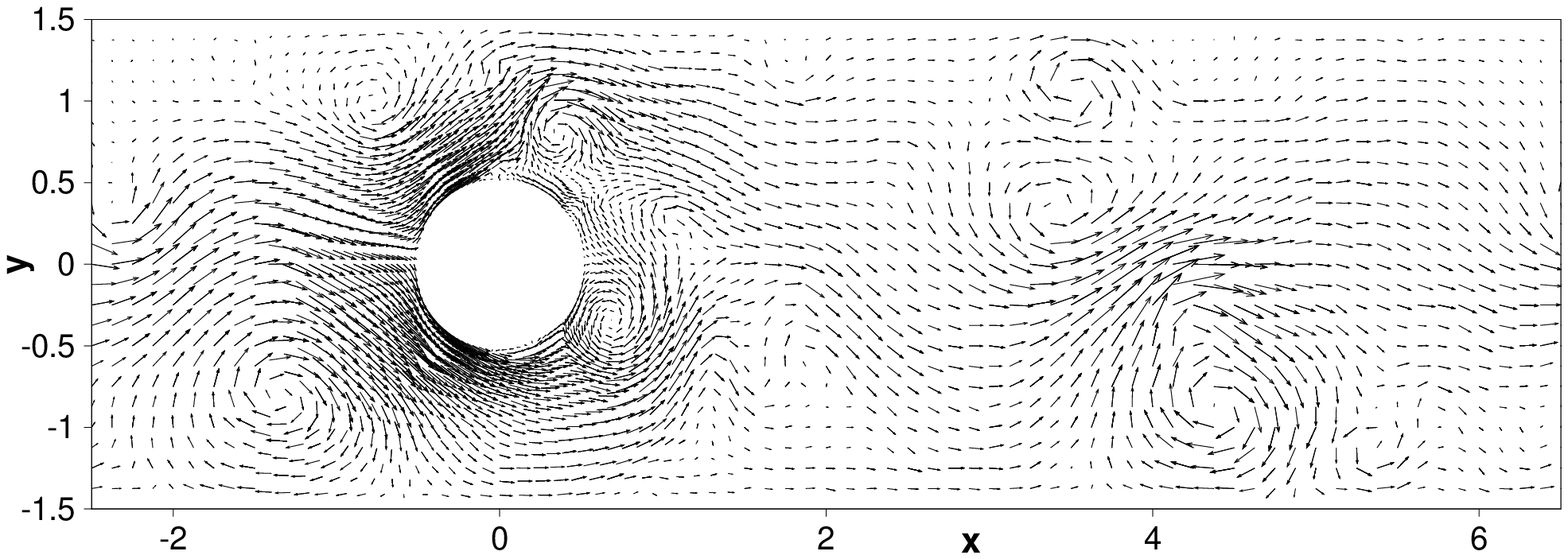}(i)
    \includegraphics[width=3.1in]{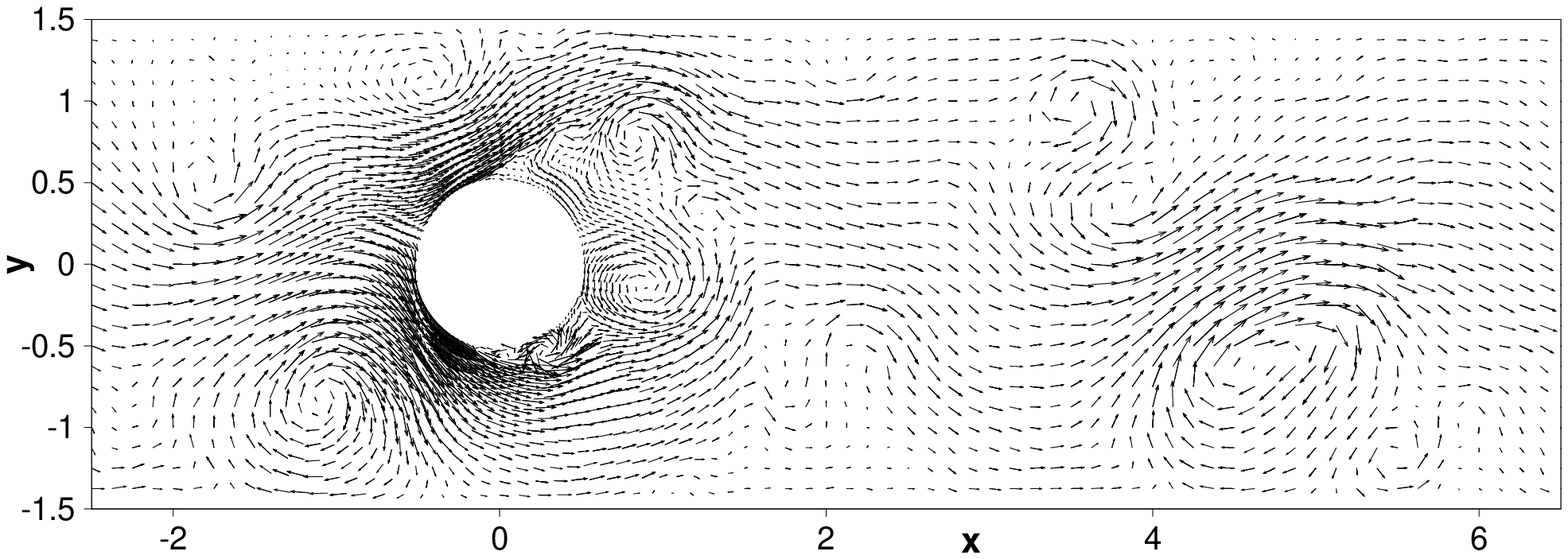}(j)
  }
  \centerline{
    \includegraphics[width=3.1in]{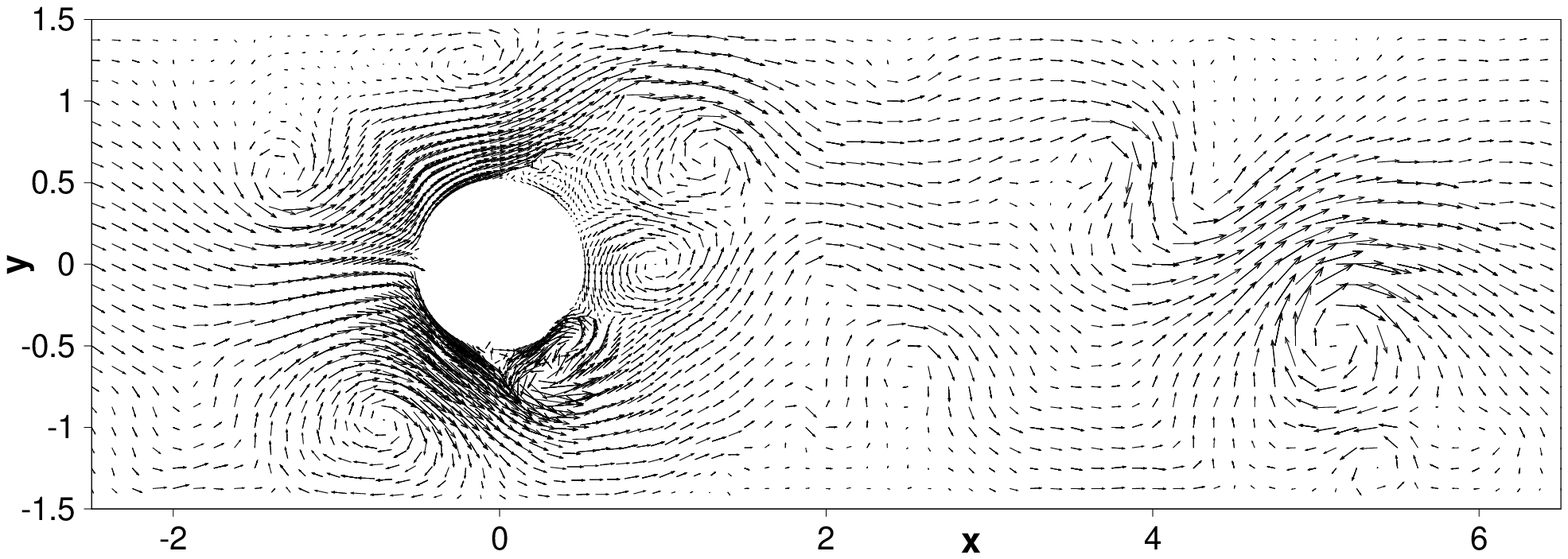}(k)
    \includegraphics[width=3.1in]{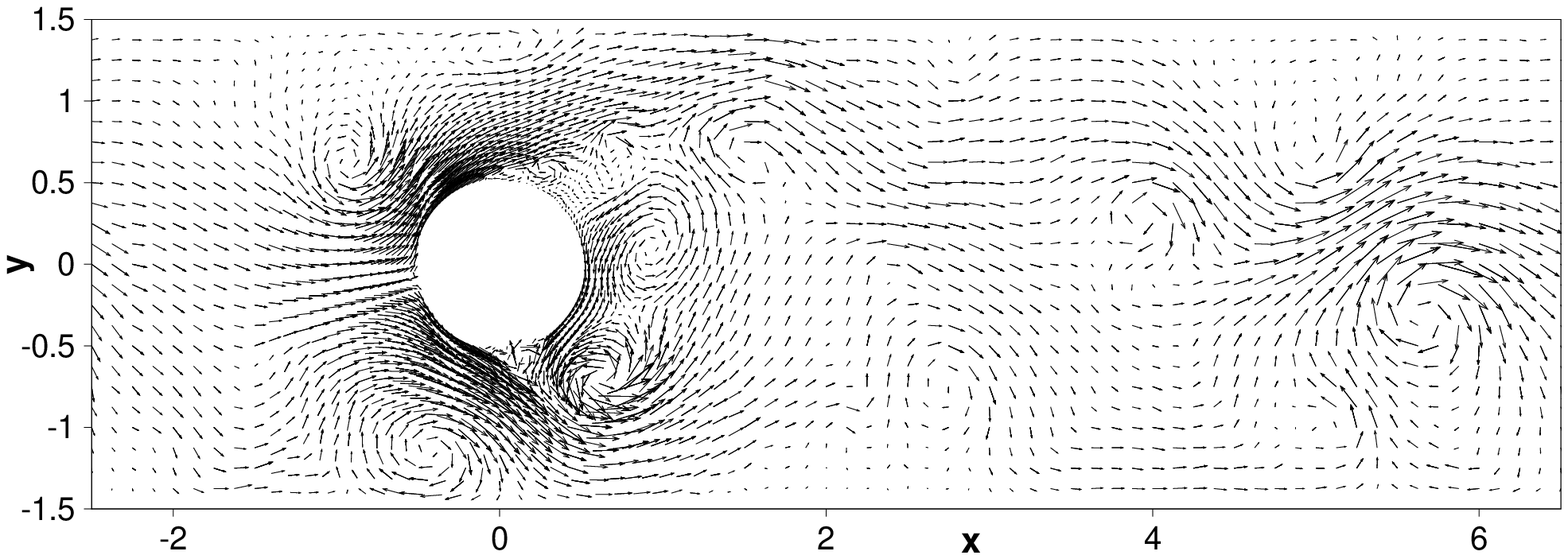}(l)
  }
  \caption{
    Temporal sequence of snapshots of the velocity fields ($\nu=0.0002$,
    total driving force $0.524$) at time
    (a) $t=3941.5$,
    (b) $t=3942$,
    (c) $t=3942.5$,
    (d) $t=3943$,
    (e) $t=3943.5$,
    (f) $t=3944$,
    (g) $t=3944.5$,
    (h) $t=3945$,
    (i) $t=3945.5$,
    (j) $t=3946$,
    (k) $t=3946.5$,
    (l) $t=3947$.
    Velocity vectors are plotted on every eighth quadrature points in each direction
    within each element.
  }
  \label{fig:vel_snap}
\end{figure}

The dynamics of this flow is illustrated by
the temporal sequence of snapshots of the velocity fields shown in
Figure \ref{fig:vel_snap} for the Reynolds number corresponding to
$\nu=0.0002$. In addition to the multitude of vortices permeating
the cylinder wake, the prominent feature of this flow
lies in the upstream vortices and their
interactions with the cylinder. Such interactions induce  complicated
dynamic features. Some upstream vortices can simply squeeze through
the gap between the cylinder and the channel wall and move downstream
into the wake, as is illustrated by the vortex marked by the symbol ``A''
in Figures \ref{fig:vel_snap}(a)-(d).
However, the interaction between the vortex and the cylinder can be more
complicated. This is illustrated by the vortex marked by ``B'' in
Figures \ref{fig:vel_snap}(a)-(g).
The incoming vortex B appears to directly collide with the
cylinder (Figures \ref{fig:vel_snap}(a)-(d)).
Upon impact, a new vortex (marked by ``C'') is spawned
near the front of the cylinder due to the strong shear layer
generated (Figure \ref{fig:vel_snap}(e)).
This causes the subsequent interactions even more dynamic.
Other  processes (such as the coalescence of vortices)
can also be observed in the wake of
the cylinder (see Figures \ref{fig:vel_snap}(h)-(j)).


\begin{table}
  \begin{center}
    \begin{tabular}{l l l}
      \hline
      & Newton-solver time/step (seconds) & Total wall time/step (seconds) \\ \hline
      Current scheme & $0.00198$ & $0.0965$ \\
      Semi-implicit scheme~\cite{Dong2015clesobc} & -- &  $0.0505$ \\
      \hline
    \end{tabular}
    \caption{
      Computational cost: comparison of wall time per time step (single processor) between
      the current scheme and the semi-implicit scheme of \cite{Dong2015clesobc}
      for the flow past a cylinder in a periodic channel with $\nu=0.0002$ (element order
      $5$). The total wall time contains the Newton solver time for the current scheme.
    }
  \label{tab:cyl_cost}
  \end{center}
\end{table}

Finally we look into the computational cost of the scheme
developed in this work. For the cylinder flow problem in a periodic channel
we have monitored the wall clock time per time step in the simulations.
In Table \ref{tab:cyl_cost} we list the typical
wall time it takes the current scheme
to compute one time step  (in seconds) on a single processor
for the Reynolds number corresponding to $\nu=0.0002$,
as well as the wall time spent
in the Newton solver for solving the scalar equation \eqref{equ:S_equ}.
These wall-time numbers are collected on a Linux cluster in
the authors' institution.
The cost of the Newton solver is insignificant,
accounting for about $2\%$ of the
total cost of the current scheme per time step.
For comparison the table also includes
the wall clock time per time step 
of the semi-implicit scheme from \cite{Dong2015clesobc}.
The current scheme requires the solution of two pressure fields
$p_1^{n+1}$ and $p_2^{n+1}$ and
two velocity fields $\mathbf{u}_1^{n+1}$ and $\mathbf{u}_2^{n+1}$,
which involves approximately twice as many operations as the semi-implicit
scheme. The computational cost of the current scheme is roughly twice
that of the semi-implicit scheme, as is evident from Table \ref{tab:cyl_cost}.


\section{Concluding Remarks}
\label{sec:summary}


We have presented an algorithm for approximating
the incompressible Navier-Stokes equations
based on an auxiliary variable associated with
the total energy of the system. This auxiliary variable is a scalar
number (not a field function), and a dynamic equation for this
variable has been introduced. This leads to a reformulated equivalent
system consisting of the modified incompressible Navier-Stokes
equations and the dynamic equation for the auxiliary variable.
The numerical scheme for the reformulated system
satisfies a discrete energy stability property in terms of
a modified energy. Within each time step, the algorithm requires
the computations in a de-coupled fashion of
(i) two pressure variables $p_1^{n+1}$ and $p_2^{n+1}$,
(ii) two velocity variables $\mathbf{u}_1^{n+1}$ and $\mathbf{u}_2^{n+1}$,
and (iii) the scalar auxiliary variable.
Computing the pressure variables and the velocity variables
involves only the usual Poisson equations and Helmholtz equations
with constant coefficient matrices.
Computing the auxiliary variable requires the solution
of a nonlinear {\em scalar} algebraic equation based on the Newton's method.
The cost for the Newton solution is insignificant and
essentially negligible (accounting for
roughly $2\%$ of the total cost per time step), because
the nonlinear equation is about a scalar number, not a field function.

The algorithm has been implemented based on a $C^0$ spectral element technique
in the current work,
and several numerical examples have been presented to test its
accuracy and performance.
The method is observed to exhibit a second-order
convergence rate in time and an exponential convergence rate in space (for smooth
field solutions). It can capture the flow field accurately when the
time step size is not too large. The method also allows the use of
large time step sizes in computations for steady flow problems,
and stable simulation results can be produced. 


The presented scheme has an attractive energy stability property, and 
it can be implemented in an efficient fashion. The algorithm
involves only constant and time-independent coefficient matrices
in the resultant linear algebraic systems, which is unlike other
energy-stable schemes for Navier-Stokes equations
(see e.g.~\cite{SimoA1994,LabovskyLMNR2009,DongS2010}, among others).
This algorithm can serve as an alternative to the semi-implicit schemes for
production simulations of incompressible flows
and flow physics studies.



\section*{Acknowledgement}
This work was partially supported by
NSF (DMS-1318820, DMS-1522537).

\bibliographystyle{plain}
\bibliography{obc,mypub,nse,sem,contact_line,interface}

\begin{thebibliography}{10}

\bibitem{AbelsGG2012}
H.~Abels, H.~Garcke, and G.~Gr\"un.
\newblock Thermodynamically consistent, frame indifferent diffuse interface
  models for incompressible two-phase flows with different densities.
\newblock {\em Mathematical Models and Methods in Applied Sciences},
  22:1150013, 2012.

\bibitem{BrownCM2001}
D.L. Brown, R.~Cortez, and M.L. Minion.
\newblock Accurate projection methods for the incompressible {N}avier-{S}tokes
  equations.
\newblock {\em J. Comput. Phys.}, 168:464--499, 2001.

\bibitem{ChenSZ2018}
H.~Chen, S.~Sun, and T.~Zhang.
\newblock Energy stability analysis of some fully discrete numerical schemes
  for incompressible navier-stokes equations on staggered grids.
\newblock {\em Journal of Scientific Computing}, 75:427--456, 2018.

\bibitem{ChenSX2012}
L.~Chen, J.~Shen, and C.J. Xu.
\newblock A unstructured nodal spectral-element method for the navier-stokes
  equations.
\newblock {\em Communications in Computational Physics}, 12:315--336, 2012.

\bibitem{Chorin1968}
A.J. Chorin.
\newblock Numerical solution of the {Navier-Stokes} equations.
\newblock {\em Math. Comput.}, 22:745--762, 1968.

\bibitem{Dong2007}
S.~Dong.
\newblock Direct numerical simulation of turbulent {T}aylor-{C}ouette flow.
\newblock {\em J. Fluid Mech.}, 587:373--393, 2007.

\bibitem{Dong2009}
S.~Dong.
\newblock Evidence for internal structures of spiral turbulence.
\newblock {\em Physical Review E}, 80:067301, 2009.

\bibitem{Dong2015clesobc}
S.~Dong.
\newblock A convective-like energy-stable open boundary condition for
  simulations of incompressible flows.
\newblock {\em Journal of Computational Physics}, 302:300--328, 2015.

\bibitem{Dong2018}
S.~Dong.
\newblock Multiphase flows of {N} immiscible incompressible fluids: a
  reduction-consistent and thermodynamically-consistent formulation and
  associated algorithm.
\newblock {\em Journal of Computational Physics}, 361:1--49, 2018.

\bibitem{DongKER2006}
S.~Dong, G.E. Karniadakis, A.~Ekmekci, and D.~Rockwell.
\newblock A combined direct numerical simulation-particle image velocimetry
  study of the turbulent near wake.
\newblock {\em J. Fluid Mech.}, 569:185--207, 2006.

\bibitem{DongS2010}
S.~Dong and J.~Shen.
\newblock An unconditionally stable rotational velocity-correction scheme for
  incompressible flows.
\newblock {\em Journal of Computational Physics}, 229:7013--7029, 2010.

\bibitem{DongS2012}
S.~Dong and J.~Shen.
\newblock A time-stepping scheme involving constant coefficient matrices for
  phase field simulations of two-phase incompressible flows with large density
  ratios.
\newblock {\em Journal of Computational Physics}, 231:5788--5804, 2012.

\bibitem{DongTK2008}
S.~Dong, G.S. Triantafyllou, and G.E. Karniadakis.
\newblock Elimination of vortex streets in bluff body flows.
\newblock {\em Phys. Rev. Lett.}, 100:204501, 2008.

\bibitem{DongZ2011}
S.~Dong and X.~Zheng.
\newblock Direct numerical simulation of spiral turbulence.
\newblock {\em J. Fluid Mech.}, 668:150--173, 2011.

\bibitem{GuermondMS2006}
J.L. Guermond, P.~Minev, and J.~Shen.
\newblock An overview of projection methods for incompressible flows.
\newblock {\em Comput. Methods Appl. Mech. Engrg.}, 195:6011--6045, 2006.

\bibitem{GuermondS2003}
J.L. Guermond and J.~Shen.
\newblock A new class of truly consistent splitting schemes for incompressible
  flows.
\newblock {\em J. Comput. Phys.}, 192:262--276, 2003.

\bibitem{HyoungsuK2011}
B.~Hyoungsu and G.E. Karniadakis.
\newblock Subiteration leads to accuracy and stability enhancements of
  semi-implicit schemes for the navier-stokes equations.
\newblock {\em Journal of Computational Physics}, 230:4384--4402, 2011.

\bibitem{JiangMRT2016}
N.~Jiang, M.~Mohebujjaman, L.G. Rebholz, and C.~Trenchea.
\newblock An optimally accurate discrete regularization for second order
  timestepping methods for navier-stokes equations.
\newblock {\em Comput. Methods Appl. Mech. Engrg.}, 310:388--405, 2016.

\bibitem{KarniadakisIO1991}
G.E. Karniadakis, M.~Israeli, and S.A. Orszag.
\newblock High-order splitting methods for the incompressible {N}avier-{S}tokes
  equations.
\newblock {\em J. Comput. Phys.}, 97:414--443, 1991.

\bibitem{KarniadakisS2005}
G.E. Karniadakis and S.J. Sherwin.
\newblock {\em Spectral/hp element methods for computational fluid dynamics,
  2nd edn.}
\newblock Oxford University Press, 2005.

\bibitem{KimM1985}
J.~Kim and P.~Moin.
\newblock Application of a fractional-step method to incompressible
  {N}avier-{S}tokes equations.
\newblock {\em J. Comput. Phys.}, 59:308--323, 1985.

\bibitem{Kovasznay1948}
L.I.G. Kovasznay.
\newblock Laminar flow behind a two-dimensional grid.
\newblock {\em Proc. Cambridge Phil. Soc.}, 44:58, 1948.

\bibitem{KravchenkoM2000}
A.G. Kravchenko and P.~Moin.
\newblock Numerical studies of flow over a circular cylinder at
  {R}e$_{D}=3900$.
\newblock {\em Physics of Fluids}, 12:403--417, 2000.

\bibitem{LabovskyLMNR2009}
A.~Labovsky, W.J. Layton, C.C. Manica, M.~Neda, and L.G. Rebholz.
\newblock The stabilized extrapolated trapezoidal finite-element method for the
  {Navier-Stokes} equations.
\newblock {\em Comput. Methods Appl. Mech. Engrg.}, 198:958--974, 2009.

\bibitem{LiuLP2007}
J.-G. Liu, J.~Liu, and R.L. Pego.
\newblock Stability and convergence of efficient {N}avier-{S}tokes solvers via
  a commutator estimate.
\newblock {\em Comm. Pure Appl. Math.}, LX:1443--1487, 2007.

\bibitem{MaKK2000}
X.~Ma, C.S. Karamanos, and G.E. Karniadakis.
\newblock Dynamics and low-dimensionality of a turbulent near wake.
\newblock {\em J. Fluid Mech.}, 410:29--65, 2000.

\bibitem{Sanderse2013}
B.~Sanderse.
\newblock Energy-conserving runge-kutta methods for the incompressible
  navier-stokes equations.
\newblock {\em J. Comput. Phys.}, 233:100--131, 2013.

\bibitem{SersonMS2016}
D.~Serson, J.R. Meneghini, and S.J. Sherwin.
\newblock Velocity-correction schemes for the incompressible navier-stokes
  equations in general coordinate systems.
\newblock {\em Journal of Computational Physics}, 316:243--254, 2016.

\bibitem{Shen1992}
J.~Shen.
\newblock On error estimate of projection methods for {N}avier-{S}tokes
  equations: first-order schemes.
\newblock {\em SIAM J. Numer. Anal.}, 29:57--77, 1992.

\bibitem{ShenXY2018}
J.~Shen, J.~Xu, and J.~Yang.
\newblock The scalar auxiliary variable (sav) approach for gradient flows.
\newblock {\em Journal of Computational Physics}, 353:407--416, 2018.

\bibitem{SherwinK1995}
S.J. Sherwin and G.E. Karniadakis.
\newblock A triangular spectral element method: applications to the
  incompressible navier-stokes equations.
\newblock {\em Comput. Meth. Appl. Mech. Engrg.}, 123:189--229, 1995.

\bibitem{SimoA1994}
J.C. Simo and F.~Armero.
\newblock Unconditional stability and long-term behavior of transient
  algorithms for the incompressible {Navier-Stokes} and {Euler} equations.
\newblock {\em Comput. Methods Appl. Mech. Engrg.}, 111:111--154, 1994.

\bibitem{Temam1969}
R.~Temam.
\newblock Sur l'approximation de la solution des equations de {Navier-Stokes}
  par la methods des pas fractionnaires ii.
\newblock {\em Arch. Ration. Mech. Anal.}, 33:377--385, 1969.

\bibitem{XuP2001}
C.J. Xu and R.~Pasquetti.
\newblock On the efficiency of semi-implicit and semi-lagrangeian spectral
  methods for the calculation of incompressible flows.
\newblock {\em International Jurnal for Numerical Methods in Fluids},
  35:319--340, 2001.

\bibitem{YueFLS2004}
P.~Yue, J.J. Feng, C.~Liu, and J.~Shen.
\newblock A diffuse-interface method for simulating two-phase flows of complex
  fluids.
\newblock {\em J. Fluid Mech.}, 515:293--317, 2004.

\bibitem{ZhengD2011}
X.~Zheng and S.~Dong.
\newblock An eigen-based high-order expansion basis for structured spectral
  elements.
\newblock {\em Journal of Computational Physics}, 230:8573--8602, 2011.

\end{thebibliography}

\end{document}